\documentclass[11pt,twocolumn]{aastex62}

\received{October 3, 2019}
\revised{April 27, 2020}
\accepted{April 28, 2020}

\shorttitle{Three-dimensional analysis of the Minispiral}

\begin{document}
\title{Three-dimensional analysis of the Minispiral at the Galactic Center:\\ orbital parameters, periods and the mass of the black hole}

\author{Maria Selina Nitschai}
\affiliation{Heidelberg University, Seminarstra{\ss}e 2, 69117 Heidelberg, Germany}
\affiliation{Max Planck Institute for Astronomy, K{\"o}nigstuhl 17, 69117 Heidelberg, Germany}

\author{Nadine Neumayer}
\affiliation{Max Planck Institute for Astronomy, K{\"o}nigstuhl 17, 69117 Heidelberg, Germany}

\author{Anja Feldmeier-Krause}
\affiliation{The Department of Astronomy and Astrophysics, The University of Chicago, 5640 S. Ellis Ave, Chicago, IL 60637, USA}

\begin{abstract}

In this paper we simultaneously fit the orbits and line-of-sight velocities of the ionized gas around the supermassive black hole, Sagittarius A$^{\ast}$ (Sgr A$^{\ast}$), at the center of the Milky Way. The data we use are taken with the K-band Multi Object Spectrograph (KMOS), presented in \cite{feldmeier15} and cover the central $\sim$ 2 pc of the Milky Way. From the brightest gas emission line in the K-band, the Br$\gamma$ line, we derive the spatial distribution and line-of-sight velocities of the gas in the minispiral. 
Using the flux distribution and the line-of-sight velocity information, we perform a fit to the three main gas streamers in the minispiral, the Northern Arm, Eastern Arm, and Western Arc, using a Bayesian modelling method, and are able to reconstruct the three-dimensional orbits of these gas streamers. With the best fit orbital parameters and the measured line-of-sight velocities, we constrain the mass of Sgr A$^{\ast}$. The orbit of the Eastern Arm is the one that is best constrained using our data. It gives a best-fit orbital period  of  $17.4_{-11.6}^{+31.0}\cdot 10^3$ years and results in an enclosed mass of $14.9_{-10.4}^{+69.4}\cdot 10^6 M_{\odot}.$

\end{abstract}

\keywords{Galaxy: center - H\textsc{ii} regions - ISM: individual objects (Sagittarius A West)- ISM: kinematics and dynamics}

\section{Introduction} \label{sec:intro}

The center of the Milky Way is by far the nearest galactic nucleus at a distance of ($8.2 \pm 0.84$)~kpc \citep{boehle16,gillessen17,abuter19}. Due to this, observations of the Milky Way nucleus yield much more detail and specific information than it would be possible for any other galactic nucleus.

Nevertheless, observations of the Galactic Center are quite challenging, since the nucleus is highly obscured by interstellar dust particles and gas in the plane of the Galactic disk, which results in magnitudes of extinction higher than A$_V \sim 30$ and reaching up to $\sim 43$ at visible wavelengths \citep{Fritz11, nogueras-lara18}. Therefore, measurements need to be carried out at longer wavelengths, in the (near-)infrared, microwave and radio bands, or at shorter wavelengths, at hard X-rays and $\gamma$-rays, where the interstellar gas and dust is more transparent.

In the inner 8 pc of the Galactic center lies the Sagittarius A complex with the radio source Sgr A$^{\ast}$, which is the central black hole of the Milky Way with a mass of $(M \sim 4.1 \pm 0.7) \cdot 10^6 M_{\odot}$ \citep{boehle16,gillessen17}.

\begin{figure}[t]
\includegraphics[scale=0.5]{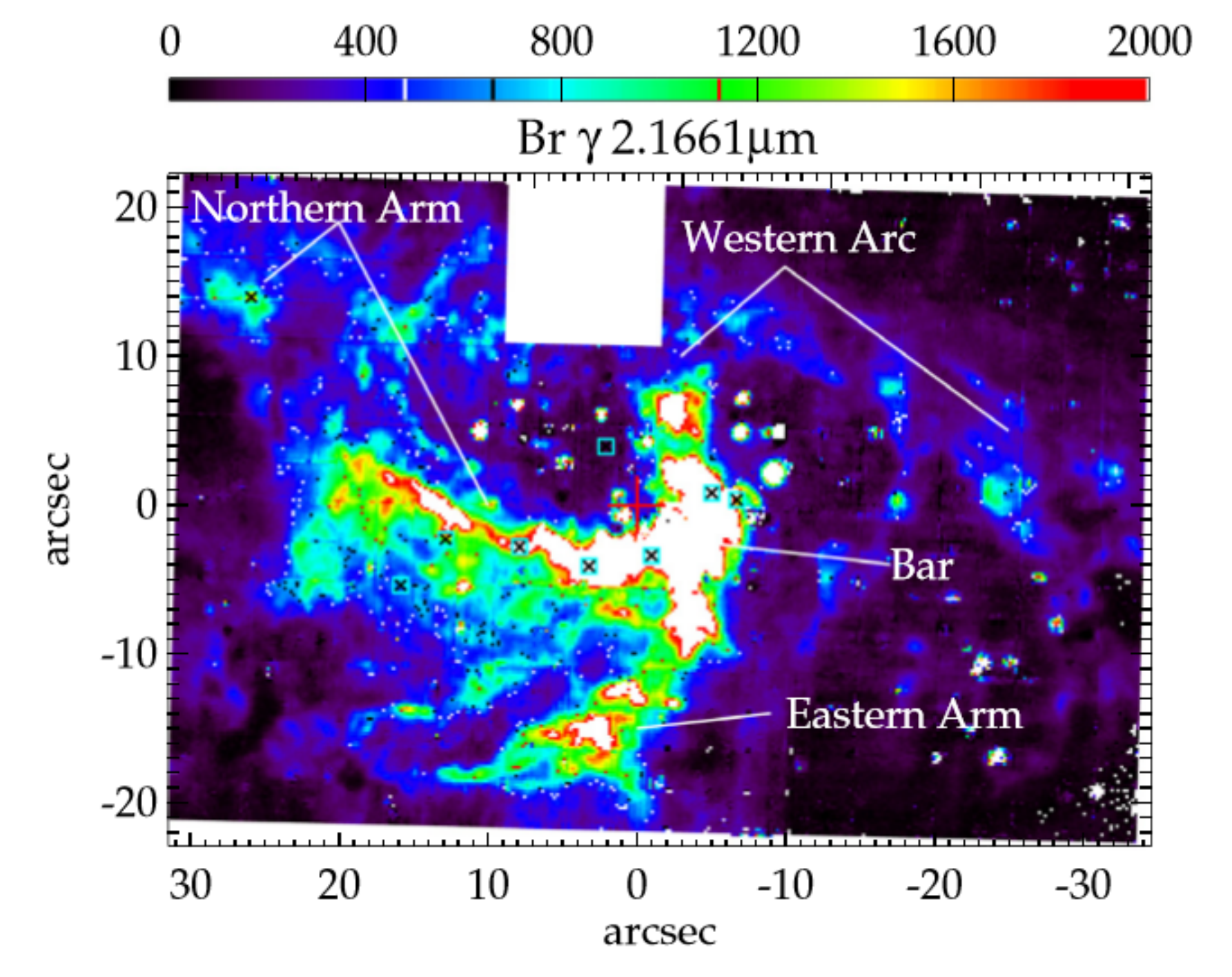}
\caption{Sgr A West emission line map of Br$\gamma$ gas of the full KMOS mosaic (white box because one arm was not working, see Section 2.3). The brightest features of the minispiral (the Northern Arm, Eastern Arm, Bar and Western Arc) are clearly visible and the red plus sign in the center indicates the position of Sgr A$^{\ast}$ {and the x-axis is aligned with the Galactic plane} 
\citep{feldmeier15}.}\label{im:Sgr A west}
\end{figure}

Surrounding Sgr A$^{\ast}$ is the circumnuclear disk (CND) that consists of a set of streamers of dense molecular gas and warm dust \citep{genzel10}. It has a ring like structure extending from 1.5 pc to $\sim$4 pc \citep{Christopher2005}. In the central region, inside the CND, lies a $\sim$ 1 to 1.5 pc radius ionized cavity with no molecular gas. This cavity consists of the H\textsc{ii} region Sgr A West and a concentration of hot, X-ray emitting gas and is pervaded by a set of orbiting ionized gas streamers, the so called minispiral or alternatively  Sgr A West \citep[see Figure~\ref{im:Sgr A west}]{Lo83, Ekers83}. The gas streamers are photoionized by combined ultraviolet radiation emitted by young massive stars in the central parsec of the Milky Way. Some of the gas streamers are orbiting mostly on circular orbits around Sgr A$^{\ast}$, but others are on elliptical orbits and get as close as 0.13 pc to Sgr A$^{\ast}$ \citep{roberts96}.

The brightest features of the minispiral are the Northern Arm, Eastern Arm, Bar and Western Arc as indicated in Figure~\ref{im:Sgr A west}.
The morphology of the minispiral results from the combination of the physical distribution of the streamers with their illumination by the ionizing radiation. The kinetic temperatures and electron densities are 5000 - 13.000 K and 3$\cdot 10^4 $ - 21$\cdot 10^4$ cm$^{-3}$ and the total mass of ionized gas in the central cavity is about 60 $M_{\odot}$ \citep{Lo83}.

The average gas density in the central parsec is thus much lower than in the surrounding circumnuclear disk, $10^{5}$ to $10^{6}$ cm$^{-3}$ \citep{Requena-Torres2012, Mills2013}, which could be a result of the young massive stars being in the post main-sequence ‘wind’ phase. The minispiral streamers represent an apparent mass inflow rate of $10^{-3}$M$_{\odot}$\,yr$^{-1}$ into the central few arcseconds \citep{genzel94}. 
The work by \cite{zhao09} and \cite{zhao10} focus on the three bright features of the minispiral describing them with Keplerian orbits, as suggested in \cite{paumard04}. The results propose that the Western Arc is in the inner ionized edge of the circumnuclear disk and has an orbit close to circular, while the Northern Arm and Eastern Arm are highly elliptical.

Even though the mass of the central black hole is better constrained with orbits of stars \citep[e.g.][]{gillessen17}, it is important to do complementary studies using the gas around the black hole. To measure the orbit of a star many measurements over a long period of time are needed while for the gas one measurement is enough, since the gas is spread along the orbit and we can constrain it with one observation using imaging spectroscopy. In addition, for other galaxies it is not possible to resolve the orbits of individual stars close to the central black hole and often gas dynamical measurements are used to constrain the black hole mass \citep{Walsh13, Davis13}. Therefore, it is essential to compare the measurements of gas and star orbits in our Galaxy and investigate their difference. Finally, to have different models and tracers that give the same result for our central black hole is also an important reassurance that we truly get the correct result.

In this paper we apply Keplerian orbits to describe the gas streamers in Sgr A West, using Br$\gamma$ observations from K-band Multi Object Spectrograph (KMOS) data \citep{feldmeier15}. 
This work follows a similar approach to \cite{zhao09}, with the difference that they used the H92$\alpha$ radio recombination line at 3.6 cm and at 1.3 cm to determine the proper motions of the H\textsc{ii} components from VLA data. The spatial resolution of their data set is superior. While our fitting method has the advantage of simultaneously including the line-of-sight velocity data to constrain the orbital parameters and not only the spatial information of the orbits. Hence, we can also include the period of the orbit as a free parameter, allowing us to constrain it without making any assumptions on the central mass. Additionally, by using a Bayesian modelling method, we get the posterior distributions that allow us to better understand the relations between the parameters and to investigate possible orbital solutions.

In Section~\ref{sec:KMOS Data} we briefly describe the observations and in Section~\ref{sec:Data Analysis} the data analysis. The results for our orbital modelling are presented in Section~\ref{sec:Orbital modelling}. Finally we discuss our results and conclude in Section~\ref{sec:Conclusion}.

Throughout this paper we use as the distance to the Galactic centre the value of $ R_{\odot} = (8178\pm13_{\mathrm{stat.}}\pm 22_{\mathrm{sys.}})$~pc \citep{abuter19} and therefore $10''\sim 0.397$~pc.

\section{KMOS Data} \label{sec:KMOS Data}

The data that we use are taken with KMOS on the ESO/VLT. The observations were performed on September 23, 2013 as part of the KMOS science verification \citep{feldmeier15}.

The 24 integral-field-units (IFUs) of KMOS were set up in a close arrangement such that an area of $64\farcs9 \times 43\farcs3$ was mapped in 16 dither positions. There is a gap in the data because one of the arms was not working (IFU13) and had to be parked during the observations (see white box in Figure~\ref{im:parts}). The total area covered corresponds to $\sim4$ pc$^2$. The map is centred on $\alpha= 266.4166^\circ$ and $\delta=-29.0082^\circ$ with a rotator offset angle at $120^\circ$, so that the long side of the mosaic is almost aligned with the Galactic plane and the area is approximately point symmetric with respect to Sgr A$^{\ast}$ \citep[RA$_{\rm Sgr A^\ast}$ = 266.41684$^\circ$ and Dec$_{\rm Sgr A^{\ast}}$ = -29.00781056$^\circ$][]{feldmeier15}. KMOS was used in the K-band ($\sim$1.934$\mu$m - 2.460$\mu$m) with a spectral resolution $R = \frac{\lambda}{\Delta\lambda}\sim4300$, which corresponds to a Full Width at Half Maximum (FWHM) of 5.55 {\AA} measured on the sky lines. The pixel scale is $\sim$ 2.8 {\AA}/pixel in the spectral direction, $0\farcs2$/pixel $\times$ $0\farcs2$/pixel in the spatial direction and the spatial resolution is 2-3 pixel measured on the FWHM of different stars in the field. The seeing during the observations varied from $0\farcs7$ to $1\farcs3$ \citep{feldmeier15} and the velocity resolution is $\Delta v$ = 75 km\,s$^{-1}$ as measured on the OH sky lines \citep{Feldmeier-Krause_2017}.

The data are presented in \cite{feldmeier15} and we refer the reader to that paper for more details on the acquisition and reduction of the data.

\section{Data Analysis} \label{sec:Data Analysis}

\subsection{Emission Line Fits} \label{subsec:Gauss Fit}
The gas streamers of the minispiral can be detected in the spectroscopic data from KMOS in the K-band through the H\textsc{i} (4-7) Brackett$\gamma$ line at 2.16613\,$\mu$m and the He\textsc{i} 2.058\,$\mu$m line. We will use only the Br$\gamma$ line since this is by far the stronger one and therefore more easily to identify in the spectra.

Emission lines have three sources in the Galactic center \citep{feldmeier15}: the extended ionized gas streamers, the molecular gas, the emission line stars.
To recognize the origin of the lines, one needs to look at their shape, since emission lines of molecular gas and stars are likely to be broader than the sharp lines of the ionized gas. Hence, we exclude all emission lines with a $\sigma$ broader than 20 \r{A} to eliminate lines and residuals from stellar or molecular gas sources.

\begin{figure}[h]
\centering
\includegraphics[width=\columnwidth]{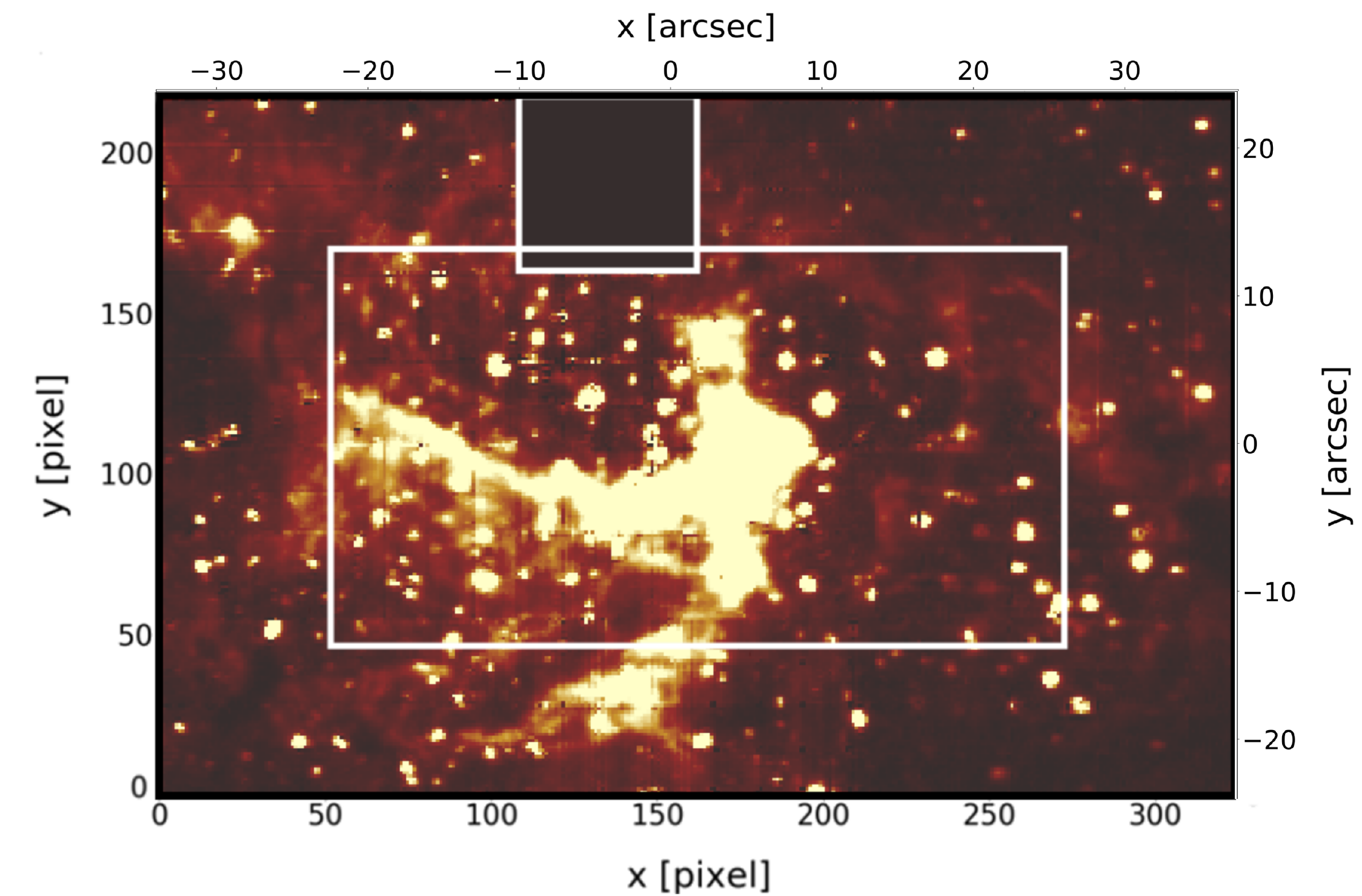}
\caption{The collapsed image of all KMOS data around the Br$\gamma$ line, from 2.156 $\mu$m to 2.171 $\mu$m, subtracting the continuum from the featureless parts between 2.065-2.107 $\mu$m and 2.177-2.207 $\mu$m. This image shows all sources of the Br$\gamma$ emission line: the ionized gas and the emission line stars. The position of Sgr A$^{\ast}$ is at the center of the image at (x$_{Sgr A^\ast}$, y$_{Sgr A^\ast}$) = (0, 0) arcsec. The inner white box is the central part of the data cube, where we fit two Gaussians and outside of the boundary we fit only one Gaussian. The small white box at the top of the image is the part where the IFU was not working.}\label{im:parts}
\end{figure}

Moreover, we split the image in two parts, the center and the boundary (see Figure~\ref{im:parts} where Sgr A$^\ast$ is at (x$_{Sgr A^{\ast}}$, y$_{Sgr A{^\ast}}$) = (0, 0) arcsec). The size of the central rectangle is chosen to include all spectra with double peaked emission lines, while in the outer part only one emission line is visible. Not every spectrum in the central box has two lines, but there is a transition between one and two lines. Hence, we allow the fit of only one Gaussian if there is only one line and two if there are two separate and comparable peaks in the center. The double peaked lines are caused by overlapping gas streamers in the two dimensional projection. At the edge the density of the ionized gas is low and therefore the Br$\gamma$ line is too weak to be fitted correctly since the continuum is as strong as the line. Therefore, in the outer part, we sum the flux over four neighbouring pixels to increase the signal in the Br$\gamma$ line.

Before fitting the emission lines, we remove a constant background in order to account for the continuum in the spectra. For each spaxel, we calculate its median value in the featureless parts between 2.065 to 2.1071 $\mu$m and 2.1772 to 2.2052 $\mu$m and we subtract this continuum level from the spectrum. For the outer part of the image this happens after we have summed over the four neighbouring pixels. 

\begin{figure}[h]
\includegraphics[width=1\columnwidth]{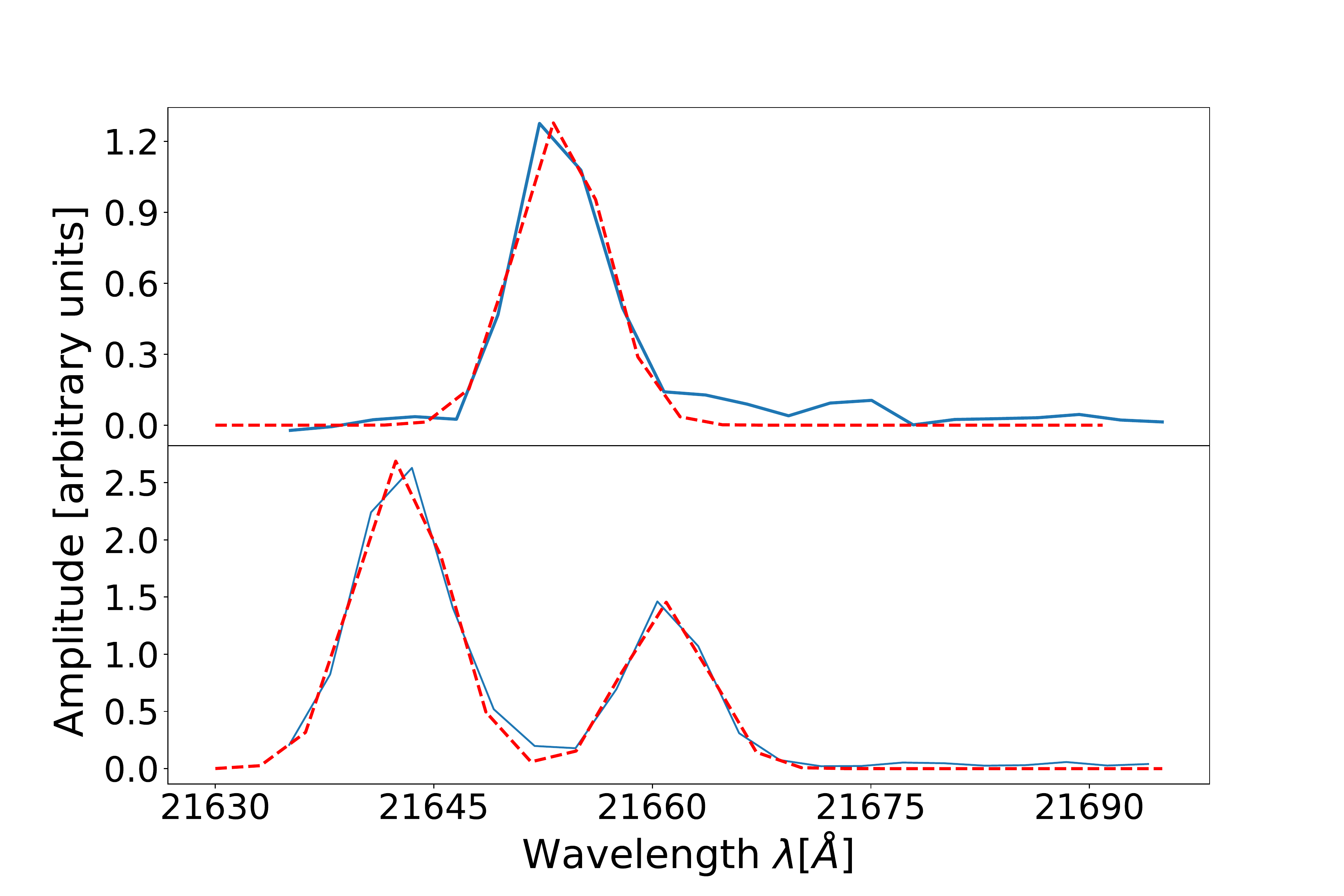}
\caption{Top panel: Single Gaussian fit to the Br$\gamma$ line. Bottom panel: Sum of two Gaussians at a double Br$\gamma$ peak. Our observed spectrum (blue) fitted by a single or double Gaussian (red dashed line).
}\label{im:Gauss}
\end{figure}

We then fit a Gaussian profile to the Br$\gamma$ line at every spaxel of the KMOS mosaic.
For the edge of the image we fit one Gaussian (see top panel of Figure~\ref{im:Gauss}) and in the central pixels, we fit a sum of two Gaussians (see bottom panel of Figure~\ref{im:Gauss}) so that we get a fit for two Br$\gamma$ emission lines at different line-of-sight velocities. 

\subsection{Velocities} \label{subsec:Velocities}
From the position of the Gaussian line fit and the width of the line we can calculate the radial velocity and the velocity dispersion of the gas using the Doppler formula. We derive the uncertainty for these values via the fit errors.

\begin{figure}[t]
\includegraphics[width=1\columnwidth]{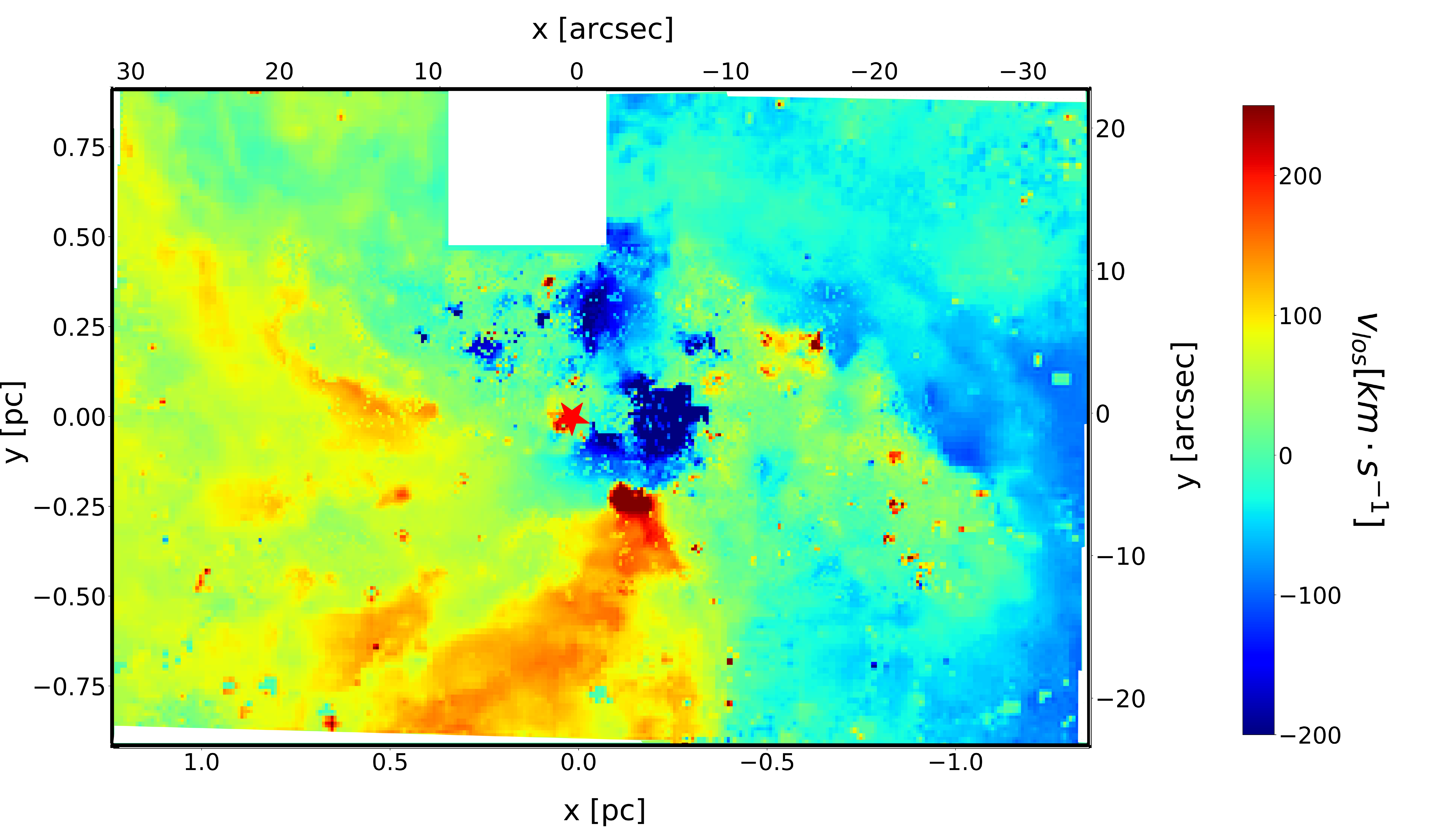}
\caption{The velocity map for the Br$\gamma$ line, as derived from the Gaussian fit to the line. In case of the double Gaussian fit, only the strongest component is used here. Several streams of fast moving gas are present. The red star indicates the position of Sgr A$^{\ast}$ and the x-axis is aligned with the Galactic plane.}\label{im:Vel}
\end{figure}

The resulting map for the velocity is shown in Figure~\ref{im:Vel}, again the position of Sgr A$^{\ast}$ is at the center of the image at (x$_{Sgr A^\ast}$, y$_{Sgr A^\ast}$) = (0, 0) arcsec = (0, 0) pc. For the pixels that were fitted with two Gaussians, only the stronger line, with the highest peak value, was used to calculate the velocity.

\begin{figure*}[t]
\centering
\includegraphics[width=1.5\columnwidth]{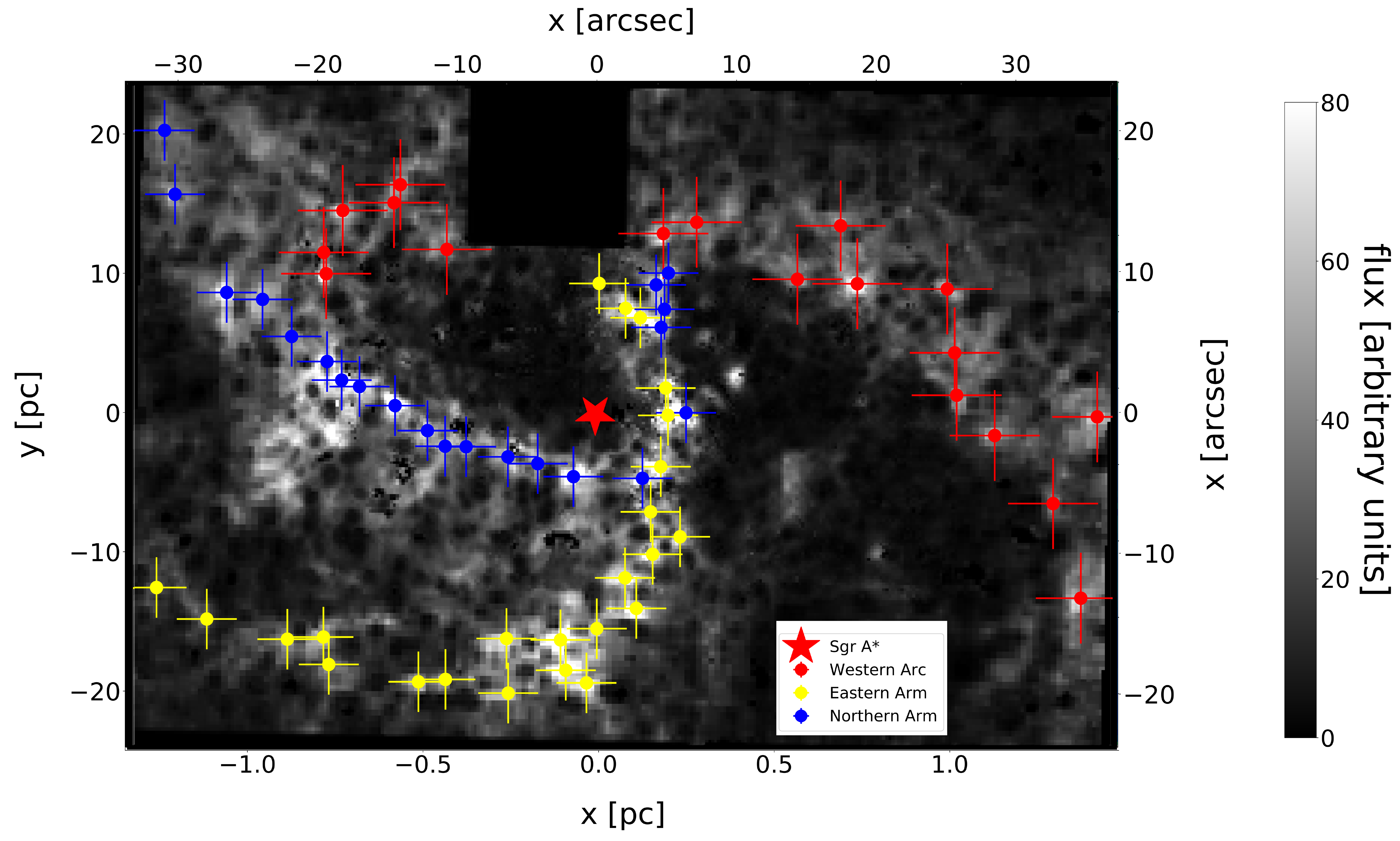}
\caption{Flux map derived from our fit to the Br$\gamma$ line, overplotted with the points we chose along the three streamers. These are the points with the highest flux in the image and their uncertainties are chosen so that they cover the surrounding area with a flux higher than 35 arbitrary flux units. Again, the x-axis is aligned with the Galactic plane}\label{im:flux punkte}
\end{figure*}

Since we have all the parameters of the Gaussian that describes the Br$\gamma$ line, we can also calculate the integral under the Gaussian, which gives us the flux of the line. This is shown in Figure~\ref{im:flux punkte}. Again, in the central region only the strongest Gaussian was used. In Figure~\ref{im:flux punkte}, one can clearly identify three structures: the Northern Arm, Eastern Arm, and Western Arc (compare to Figure~\ref{im:Sgr A west}).

The range of radial velocities is from $\sim -350$ km\,s$^{-1}$ to 350 km\,s$^{-1}$ over the entire field of view and the median error is $\sim$3.74 km\,s$^{-1}$. In the three main features of Sgr A West, the Northern Arm, Eastern Arm and Western Arc, the median velocity is 69.87 km\,s$^{-1}$. The velocities agree with other measurements \citep[e.g.][]{wollman77,roberts96,paumard04,zhao09}.

\section{Orbital modelling} \label{sec:Orbital modelling}

\subsection{Bayesian Fit} \label{subsec:Max fit}

Once we have extracted all the information about the velocity from the data, we now look at the motion of the gas streamers. To do this, we need to find some points in the data that can clearly be identified as points on the three main streamers (Northern Arm, Eastern Arm and Western Arc). Therefore, we use the flux data from the previous section and look at the areas with the highest flux. The points we selected are marked on the flux map (Figure~\ref{im:flux punkte}). These points are the spatial pixels with the highest flux. Their spatial uncertainties are chosen so that they cover at least the surrounding area with a flux higher than $\sim$35 arbitrary flux units (typically 10 pixels, i.e. 0.0793~pc). However, for the Western Arc a higher uncertainty of 15 pixel (0.119 pc), was chosen since it is more diffuse and broader than the other two streamers. To decide to which streamer each point belongs, especially for the Northern and Eastern Arm in the central region, we use the higher resolution work of \cite{zhao09} as orientation. They had identified H\textsc{ii} components belonging to certain streamers by using also their proper motion. For each selected point on the flux map we use the radial velocity calculated at that spatial pixel together with its error from the Gaussian fit.  

The gas streamers can be described with Keplerian orbits \citep{paumard04,zhao09}, which means that the gas moves along an ellipse with the central black hole at one focus. To describe the geometry of the orbits we need five parameters for the orbital elements for each streamer: 1) the semimajor axis (\textit{a}), 2) the eccentricity (\textit{e}), 3) the inclination (\textit{i}), 4) the longitude of the ascending node (\textit{$\Omega$}), 5) the argument of the perihelion ($\omega$). In addition we need the period (\textit{T}) along the orbit to calculate the orbital velocity at each point of the orbit which enables us to fit the velocity component projected along the line-of-sight. Moreover, we have allowed for two extra parameters in the fit in order to account for the pointing uncertainties of KMOS. These, are $f_x$ and $f_y$ that can have values from $\pm$ 10 pixel, but we expect them to have values around 0. They are added to the position of Sgr A$^{\ast}$ before centering the data on that position.

To fit Keplerian orbits to these selected points, we use the Kepler Ellipse class from the Python package PyAstronomy \footnote[4]{https://github.com/sczesla/PyAstronomy}. This class requires six parameters (\textit{a, T, e, i, $\Omega$, $\omega$}) to plot the orbit and to describe all the orbital features.

The formal statistical errors of the velocity data are quite small and certainly much smaller than the actual deviations. To approximately account for systematic uncertainties of the data we follow the approach of Sec.~3.2 of \citet{vandenBosch2009}, as modified for Bayesian analysis in Sec.~6.1 of \citet{mitzkus17}. In brief, we scale the errors to reach a meaningful $\Delta\chi^2$ difference that is equivalent to the standard deviation of the $\chi^2$, which is $\Delta\chi^2=\sqrt{2N}$, for N data points. To include this uncertainty in the MCMC sampling, we increase the errors by multiplying all errors $\epsilon$ by $(2N)^{1/4}$. This is equivalent to changing the $1\sigma$ confidence level from $\Delta\chi^2=1$ to $\Delta\chi^2=\sqrt{2N}$.
To do this, we initially adopt a constant error for the kinematics of the Northern Arm, the Eastern Arm and the Western Arc, of $\epsilon=(5.3, 3.5, 6.8)$ km\,s$^{-1}$ respectively, to give $\chi^2/DOF = 1$. Then we multiply $\epsilon$ by $(2N)^{1/4}$ as described above. This is equivalent to redefining the $\Delta \chi^2$ confidence level, taking the $\chi^2$ uncertainty into account.

The Bayesian modelling was performed using the \texttt{adamet} package\footnote[5]{We use the Python version 2.0.7 of the \texttt{adamet} package available from  https://pypi.org/project/adamet/} of \citet{cappellari13}, which implements the Adaptive Metropolis algorithm of \citet{haario01}. This is used to estimate in an efficient way the posterior distribution, as in standard MCMC methods, to get the confidence levels of the best fitting parameters and to show the relations between the different parameters. We adopted uniform priors on all parameters, in such a way that the probability of the model, given the data, is just the likelihood $P({\rm data}|{\rm model})\propto\exp(-\chi^2/2)$. 

\subsection{Orbital Parameters} \label{subsec:orbit_params}

\begin{figure*}[ht]
 \begin{minipage}{0.5\textwidth}
 \includegraphics[width=1\columnwidth]{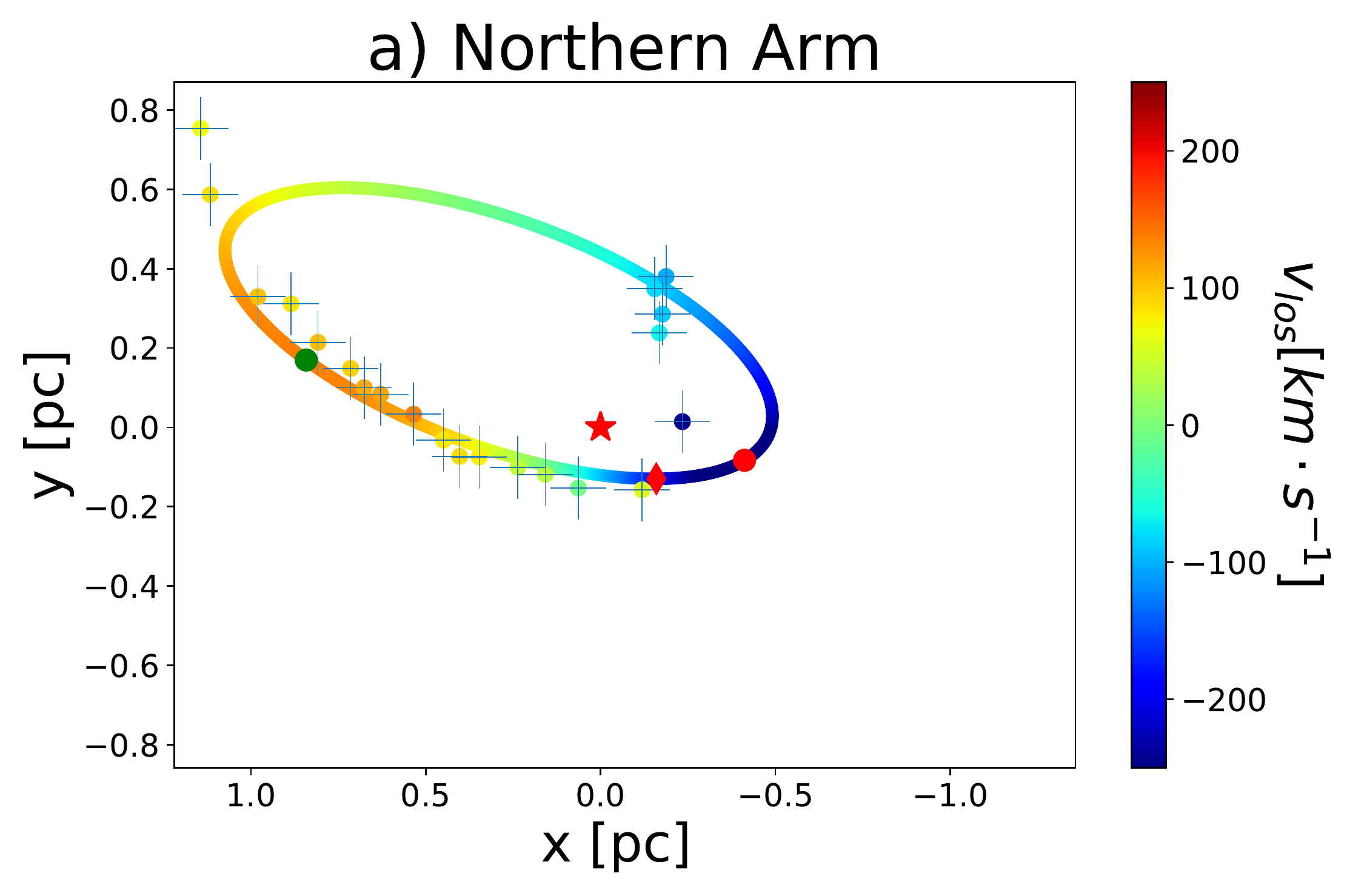}
 \end{minipage}
 \begin{minipage}{0.5\textwidth}
\includegraphics[width=1\columnwidth]{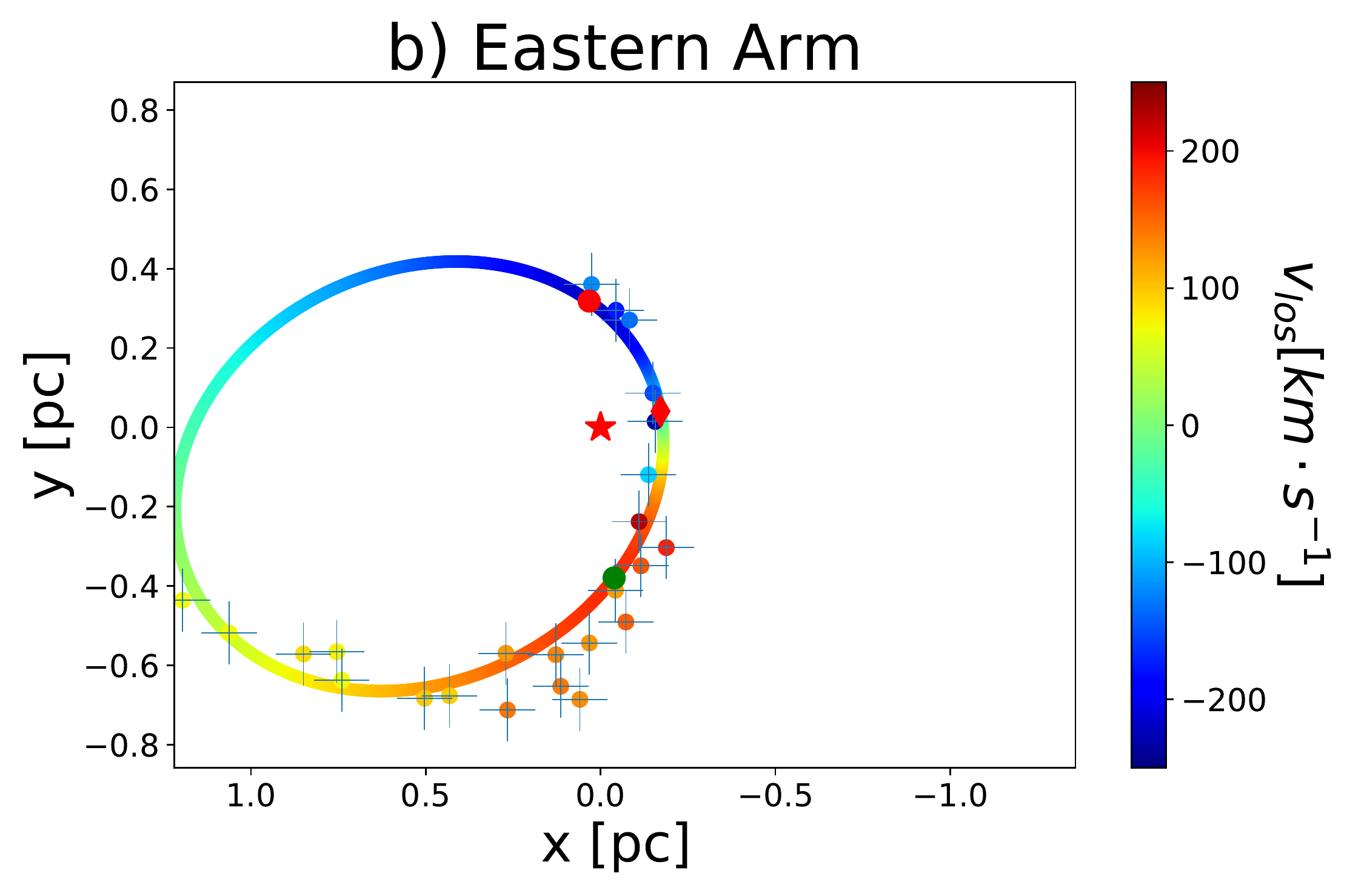}
\end{minipage}
\begin{minipage}{0.5\textwidth}
\centering
\includegraphics[width=1\columnwidth]{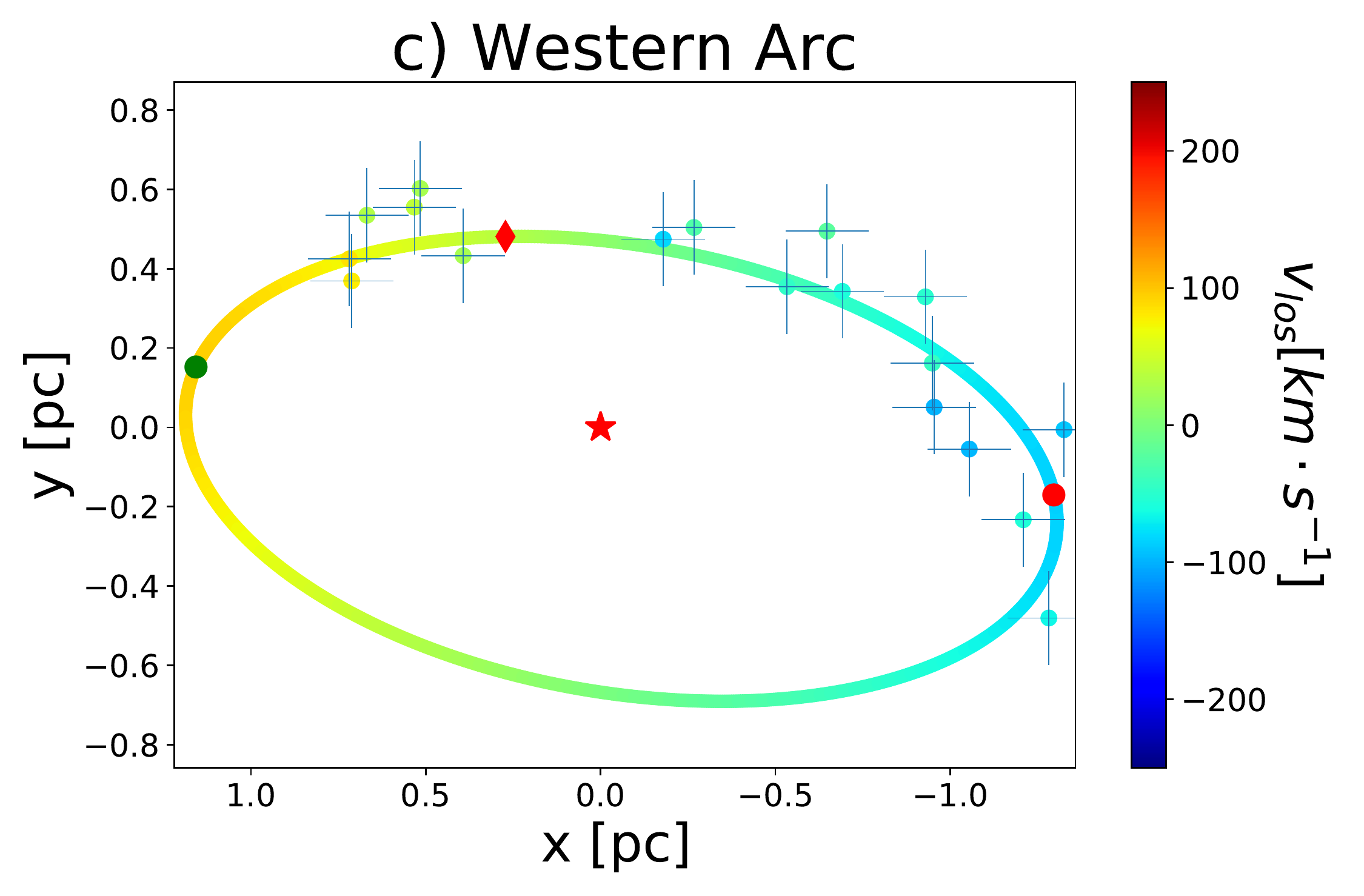}
\end{minipage}
\caption{The best-fit Keplerian orbits for the Northern Arm, Eastern Arm and Western Arc with the projected line-of-sight velocities along the orbit. The points of the fitted orbit are colour coded with their respective line-of-sight velocities. The velocities of the data are the observed line-of-sight velocities of Section~\ref{subsec:Velocities}. The red star indicates the position of Sgr A$^{\ast}$, the red diamond is the periapsis, the green circle is the ascending node, the red circle is the descending node and again the x-axis is aligned with the Galactic plane.}\label{im:vel}
\end{figure*}

The five geometric parameters set the shape of the orbit, while the period \textit{T} provides the orbital velocity at each point, and we fit the projected geometry and the line-of-sight velocities simultaneously.
The results for our fits are shown in Figure~\ref{im:vel}, here, the line-of-sight velocities along the orbit are given by the plotted colour. The best-fit model parameters are listed in Table~\ref{table:parameters}. The probability distributions for our best fit results are shown in the Appendix, in Figures~\ref{im:MCMC n},~~\ref{im:MCMC e}, and~\ref{im:MCMC w} for the three different streamers. In case there are more than one solution the 1$\sigma$ errors in Table~\ref{table:parameters} are calculated applying a cut to only use the posterior distribution around the best-fitting solution ($\omega_{\mathrm North} < 180^\circ$ and $\omega_{\mathrm West} > 200^\circ$).

\begin{table*}[ht]
\centering
\small
\caption{Best fit orbital parameters of the Keplerian orbits, and inferred black hole mass}\label{table:parameters}
\begin{tabular}{lccr}
\toprule
 Best Fit Parameters & Northern Arm & Eastern Arm & Western Arc\\
\hline
Eccentricity, \textit{e}&  $ 0.65^{+0.11}_{-0.15}$ & $0.752_{-0.099}^{+0.092}$ &  $ 0.18_{-0.13}^{+0.30}$\\[1.2ex]
\hline
log(Semimajor axis), log(\textit{a}) [log(arcsec)] & $ 1.388^{+0.085}_{-0.076}$ & $ 1.305_{-0.077}^{+0.127}$ & $ 1.51_{-0.09}^{+0.17}$\\[1.2ex]

Semimajor axis, \textit{a} [arcsec] & $ 24.5_{-4.0}^{+4.9}$ & $ 20.2_{-3.3}^{+7.3}$ & $ 32.1_{-6.4}^{+15.2}$\\[1.2ex]

Semimajor axis, \textit{a} [pc] & $ 0.97_{-0.16}^{+0.20}$ & $ 0.80_{-0.13}^{+0.29}$ & $ 1.27_{-0.26}^{+0.60}$\\[1.2ex]
\hline
Longitude of the ascending node, $\Omega$ [deg] & $ 281.4^{+6.5}_{-6.4}$& $ 174.2_{-8.5}^{+8.6}$ & $ 277.5_{-8.5}^{+7.7}$\\[1.2ex]
\hline
Argument of the perihelion, $\omega$ [deg] & $ 122.0^{+9.3}_{-7.5}$ & $ 96.7_{-7.0}^{+9.7}$ &   $ 288.5_{-23.7}^{+42.3}$\\[1.2ex]
\hline
Inclination, \textit{i} [deg]& $ 109.2^{+7.9}_{-6.8}$ & $ 152.2_{-20.9}^{+17.1}$&   $ 116.4_{-49.3}^{+25.9}$\\[1.2ex]
\hline
Uncertainty x-position Sgr A$^{\ast}$, $f_x$ [pix]& $ 0.82^{+7.21}_{-7.68}$ & $ 4.6_{-6.9}^{+4.0}$ &   $ 0.15_{-7.1}^{+7.0}$\\[1.2ex]
\hline
Uncertainty y-position Sgr A$^{\ast}$, $f_y$ [pix] & $-1.23^{+7.28}_{-6.43}$ & $-2.1_{-5.7}^{+7.2}$ &   $-0.05_{-6.7}^{+6.8}$\\[1.2ex]
\hline
log(Period), log(\textit{T}$_{\mathrm{fit}}$) [log($10^3$ yr)] & $1.55^{+0.15}_{-0.12}$ & $1.24_{-0.47}^{+0.42}$ &   $1.90_{-0.13}^{+0.15}$\\[1.2ex]

Period, \textit{T}$_{\mathrm{fit}}$ [$10^3$ yr] &$ 35.3_{-8.5}^{+14.0}$ & $ 17.4_{-11.6}^{+31.0}$ & $ 79.9_{-20.8}^{+32.2}$\\[1.2ex]
\hline
Black hole mass, M$_{\mathrm{BH}}$ [$10^6$ M$_{\odot}$] & $ 6.4_{-2.5}^{+3.9}$ & $ 14.9_{-10.4}^{+69.4}$ & $ 2.8_{-1.0}^{+3.3}$\\[1.2ex]
\toprule
\end{tabular}
\end{table*}

\begin{figure*}[ht]
 \begin{minipage}{0.5\textwidth}
 \includegraphics[width=1\columnwidth]{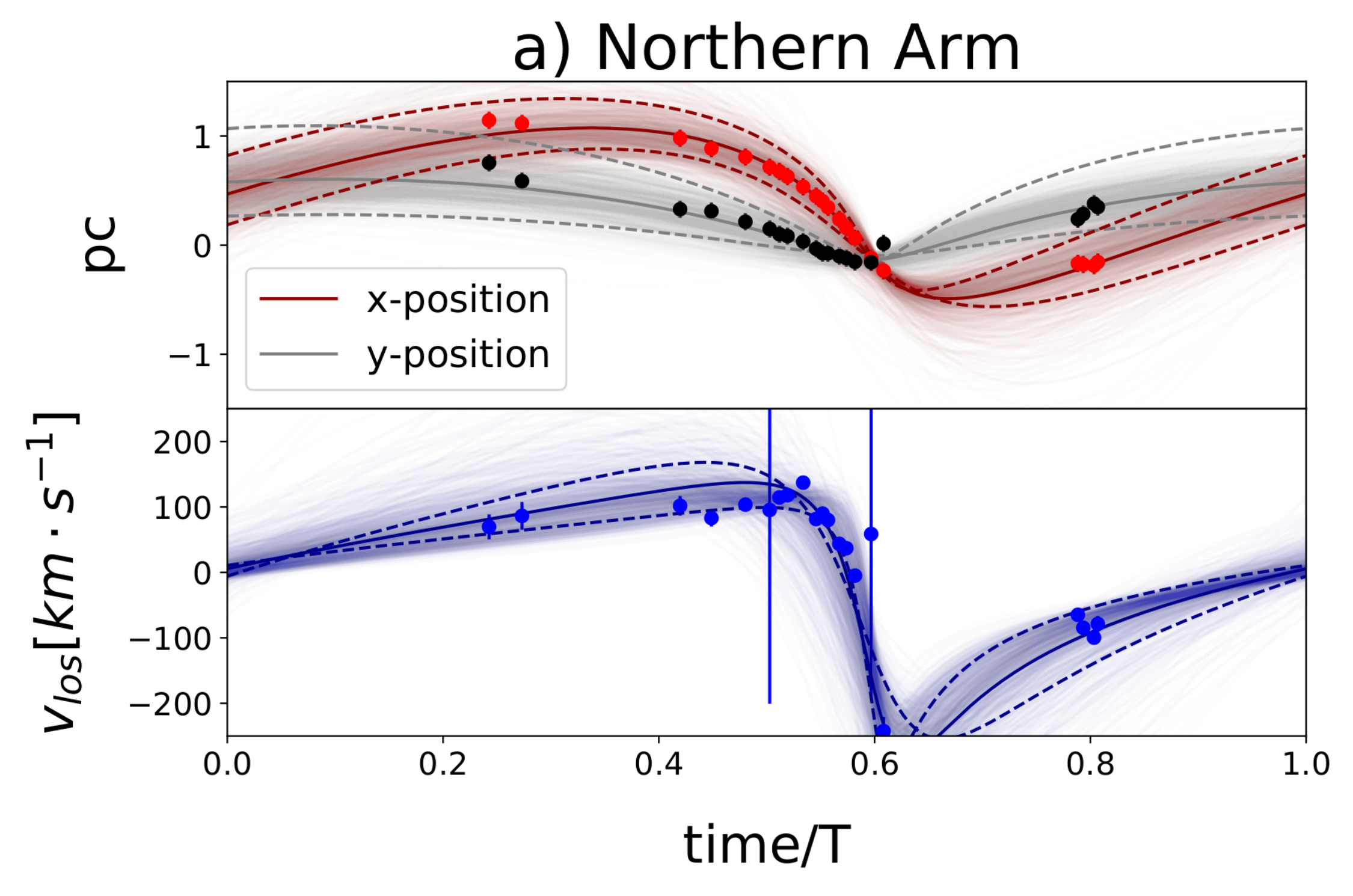}
 \end{minipage}
 \begin{minipage}{0.5\textwidth}
\includegraphics[width=1\columnwidth]{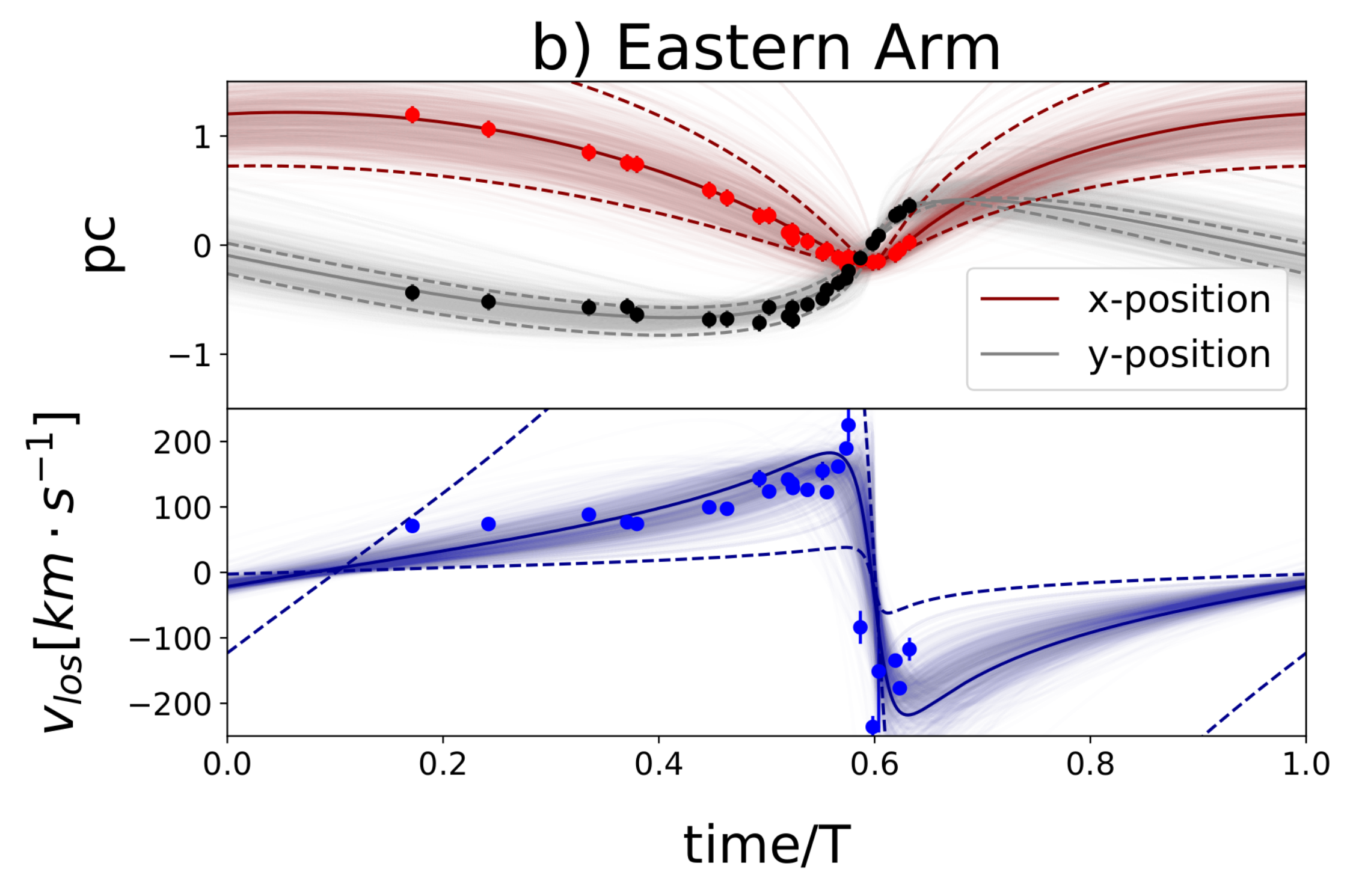}
\end{minipage}
\begin{minipage}{0.5\textwidth}
\centering
\includegraphics[width=1\columnwidth]{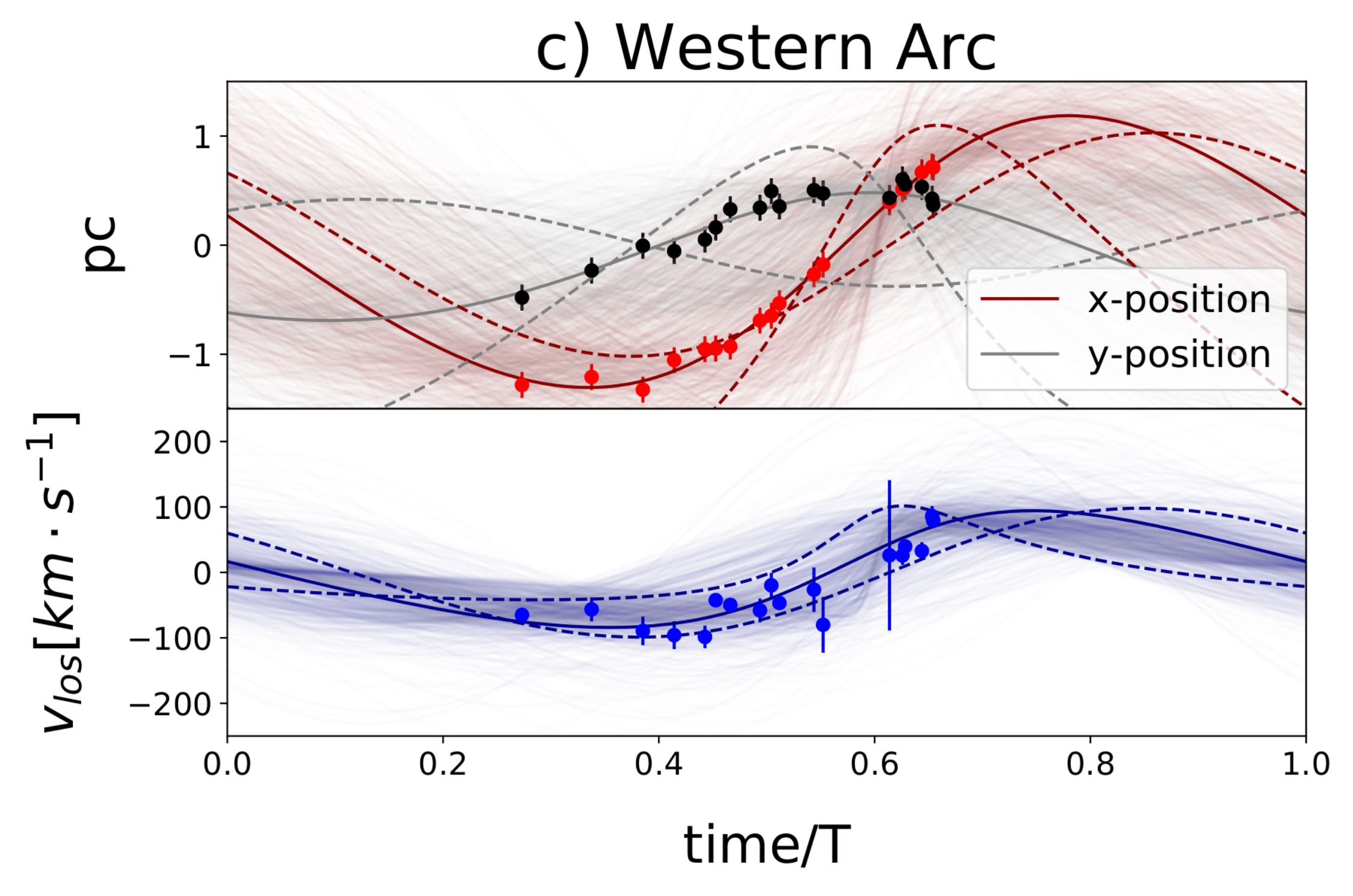}
\end{minipage}
\caption{The x, y position in the plane of the sky and the velocity versus the time for the three streamers. The data points are plotted as dots and the best-fit orbit as solid line, while the random orbits from the posterior are transparent lines and the dashed lines are the  1$\sigma$ ranges. The x position has the colour red, the y position the colour black and the velocity is plotted in blue. The time on the $x$-axis of the plot is normalized to the period of the orbits. For the Northern Arm the posterior used for the transparent lines fulfill the condition $\omega < 180^\circ$ and for the Western Arc $\omega > 200^\circ$.}\label{im:time}
\end{figure*}

In addition, in Figure~\ref{im:time} we plot the x, y position and the velocity versus time for the data points together with their errors and the best model together with a random set of the possible orbits from our posteriors. The Eastern Arm has one solution only; this is directly visible from the posterior distribution in Figure~\ref{im:MCMC e}. Looking at the line-of-sight velocities, the position of the periapsis as well as the ascending and descending nodes are very well constrained on the orbit, see Figure~\ref{im:vel}, as they fall where the line-of-sight velocity changes sign. For the Northern Arm and Western Arc there are two solutions for the inclination and argument of perihelion that would fit the data (while turning the orbit), see posterior distributions in Figures ~\ref{im:MCMC n} and ~\ref{im:MCMC w}. Especially for the Western Arc the orbit could be turned around by $\sim 180^{\circ}$ while still fitting the data, since it is not so clear where the position of the periapsis as well as the ascending and descending nodes are. Hence, we have added the condition, $\omega_{\mathrm North} < 180^\circ$ and $\omega_{\mathrm West} > 200^\circ$ accordingly, for the random selection from the posterior for Figure~\ref{im:time} a) and c) in order to just choose posteriors around the best solution and not close to the second possible solution. One can clearly see that the Western Arc is not as well constrained as the other two orbits since the scattering of the random orbits from the posterior is much broader around the data points than for the other gas streamers. In addition, the data points we mainly have for the Western Arc are on one side between the ascending and descending node and not towards the other side like for the other streamers. This fact allows for some freedom for the orbital parameters since it does not allow us to constrain the shape of the orbit as well as for the other streamers. However, the best-fitting solution agrees well with the data points. In addition, we have checked also the orbits around the second solutions for the Northern Arm and Western Arc. These solutions would give similar orbits to Figure~\ref{im:vel} but the time plots of Figure~\ref{im:time} would have an offset of 180$^\circ$ between the data and the models for $x$ and $v_{\mathrm los}$. Thus, Figure~\ref{im:time} reassures us that the best-fitting values are the solutions with the highest likelihood for all streamers (the red dashed line in the posterior distributions Figures~\ref{im:MCMC n},~~\ref{im:MCMC e} and~\ref{im:MCMC w}).

Moreover, the inclination angle of all three streamers is larger than 90$^\circ$. This means that the orbital motion is counterclockwise as viewed from the Earth \citep{karttunen}.

We also see that the Western Arc is more circular (e = 0.18$_{-0.13}^{+0.30}$) than the other orbits as it has been suggested in \cite{roberts93} and the other two streamers are highly elliptical \citep{zhao09}.

\begin{figure*}
    \centering
    \includegraphics[width=1.5\columnwidth]{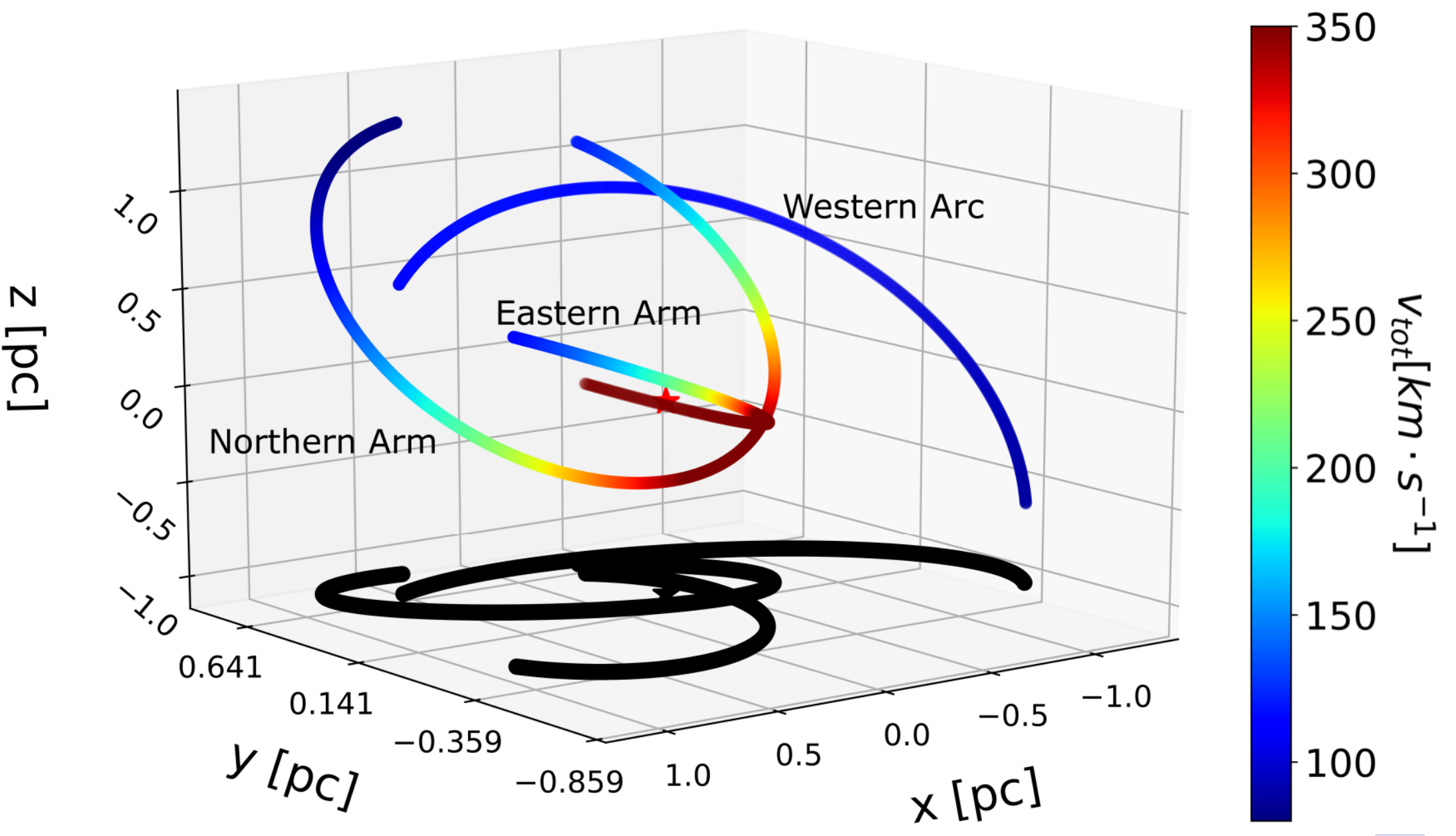}
    \caption{The three dimensional orbit prediction for the three streamers from our best-fit. Here, the colour coding represents their total three dimensional orbital velocities. The positive z-direction is the direction to the observer. The gas in the Eastern arm reaches orbital velocities of up to 350 km s$^{-1}$ while for the Western Arc the velocities are below 150 km s$^{-1}$. The black orbits are the projection on the plane of the sky of the three dimensional gas streamers.}
    \label{im:total plot}
\end{figure*}

In Figure~\ref{im:total plot} all three streams are shown in three dimensions with their three dimensional orbital velocity. The Western Arc has by far the lowest orbital velocity since it is the furthest away from the black hole, while the Eastern Arm has the highest orbital velocities close to the center. This plot is consistent with the work of \citet[Fig. 5]{zhao10} for the Northern and Eastern Arm but differs for the Western Arc. However, if we chose the second solution ($i = 65.8^{+46.8}_{-26.3}$, $\omega= 74.6^{+28.3}_{-42.5}$) for the Western Arc from the posterior, the figure is the same as the one in \cite{zhao10}. This means that our second solution is agreeing better with the previous results from \cite{zhao09, zhao10} but the first solution has the highest likelihood and the time plot (see Figure~\ref{im:time}) agrees better with the data points. 

If we compare our results with the work of \cite{zhao09} for the  H92$\alpha$ radio line, we can see that our parameters overall agree with their findings (see Table 5 in \citealt{zhao09}). The eccentricity and semimajor axis value all agree within their error range and the inclination and $\omega$ agrees in the 3$\sigma$ uncertainties. For the Western Arc, the value for $\omega$ from Table~\ref{table:parameters} does not agree with the value from \cite{zhao09} but the second solution agrees within the error range. This is consistent with what we already show with Figure~\ref{im:total plot} and hence it seems that \cite{zhao09} has just selected the second solution but otherwise we agree with their findings. In addition, we agree with \cite{zhao09, paumard04} that show that the Northern and Eastern Arms are coming close to the central black hole and that the Western Arc is the furthest away from it. However, the $\Omega$ values do not agree with the values of \cite{zhao09}, but are offset by $\sim 180^\circ$ probably because we have set a different 0$^\circ$ position for it.

Moreover, from the posterior distributions in Figure~\ref{im:MCMC n}, \ref{im:MCMC e} and \ref{im:MCMC w} we also can see that $f_x$ and $f_y$ are not tightly constrained, especially for the Western Arc were they could get any value. The Eastern Arm seems to prefer higher values for  $f_x$  and lower values for  $f_y$, but both $f_x$ and $f_y$ are in agreement with 0 pixel.

All selected points with their velocities and uncertainties for each streamer are listed in Table~\ref{table:data} in the Appendix after adding the best fitting $f_x$ and $f_y$ for the position of Sgr A$^{\ast}$.

\subsection{Mass Calculation} \label{subsec:Mass calculation}
Since we have derived all the required parameters for the best fit of the Kepler orbits for the three streamers, we can now calculate the mass of the black hole, Sgr A$^{\ast}$.

In general, for any Keplerian orbit, Kepler's third law gives a relation between period and mass:
\begin{equation}\label{eq:period}
T = 2\pi\sqrt{\frac{a^3}{M\,G}},
\end{equation}
where M is the enclosed mass and $G = 6.674\cdot10^{-11}\mathrm{m}^3\,\mathrm{kg}^{-1}\,\mathrm{s}^{-2}$ is the gravitational constant.

Using the above equation we can therefore derive the enclosed mass which is comparable to the mass of the central black hole \citep[e.g.][]{Feldmeier-Krause2017} from the streamer's fitted orbits. The results are ($ 6.4_{-2.5}^{+3.9}$, $ 14.9_{-10.4}^{+69.4}$, $ 2.8_{-1.0}^{+3.3}$) $\cdot$ $10^6 M_{\odot}$ for the Northern Arm, Eastern Arm and Western Arc, respectively and are listed in Table~\ref{table:parameters}. The uncertainties are the 1$\sigma$ range calculated as the 68 per cent of the same random posterior values used Figure~\ref{im:time}. In the same way the uncertainties for the Period in years and the semimajor axis in arcsec or pc are derived.

Additionally, if we compare our results of the black hole mass with recent measurements, like $(4.28 \pm 0.10_{stat}\pm 0.14_{syst})\cdot 10^6 M_{\odot}$ \citep{gillessen17}, $(4.1\pm 0.6) \cdot 10^6 M_{\odot}$ \citep{ghez08} and 
$(4.02\pm0.16_{stat}\pm0.04_{syst})\cdot 10^6 M_{\odot}$ \citep{boehle16}, we see that our results of the mass (see Table~\ref{table:parameters}) agree within the error range of our measurements.

The Eastern Arm, however, over predicts the black hole mass. From the posterior (see Figure~\ref{im:MCMC e}) one sees that the possible periods allowed are reaching lower values compared to the other streamers and hence give higher mass estimates. However, we obtain reasonable values for the mass if we consider higher periods that are still in the 1$\sigma$ range. The Western Arc is the only streamer that obtains a best-fit value that is lower rather than higher than the literature values but still agreeing in the 1$\sigma$ range. This could be explained by the fact that this streamer is rather diffuse, which makes it difficult to select exact orbit points. Unlike the other streamers, for which we can easily identify high flux points that lie on the orbit, the Western Arc is rather a broad band, with similar low flux in more extended regions. Moreover, like already mentioned in the previous Section, Figure~\ref{im:time} indicates that the Western Arc is not well constrained since  there is a large variety of possible orbits that describe the data reasonably well. 

Also, the fitted enclosed mass, using a similar technique, taken from \cite{zhao09} of $4.2_{-3}^{+6}\cdot 10^6 M_{\odot}$ is in agreement with our results and the periods for their orbits (except the Eastern Arm) are within the same range as ours, $\sim$(4-8)$\cdot$10$^4$ yr (see Table 5 in \cite{zhao09}). The period value is also in agreement with the earlier work of \cite{paumard04}, where they use the He\textsc{ i} and the Br$\gamma$ line but with poorer resolution than our data and give for the Northern Arm a period between a few 10$^4$ to a few 10$^5$ yr.

Since for our calculation for the enclosed mass only the semimajor axis and period are needed, the two solutions for the Northern Arm and Western Arc do not influence our results. The $a$ and $T$ are not changing significantly for the second solutions and only the angle values have two clear separate solutions seen in the posterior distribution.

In addition, we tested how much our results change if we select differently the overlapping points for the Northern and Eastern Arm. First, we selected the six points that are close to the center and for which it is not obvious to which streamer they belong. Then we redid our fits with either these points exchanged between the two streamers, or including all six points to each streamer. For both, the Northern and Eastern Arm the results do not change significantly, and they all agree within the 1$\sigma$ uncertainties. We investigated the fit residuals for the Eastern Arm and found that the new orbit fits do not fit the data as well as our original best fit. The velocities of the central data points, which we exchanged or included in this test, are too low to match to the velocities of the rest of the Eastern Arm. For the Northern Arm, on the other hand, the velocities of the points we exchanged are too high. We conclude that the central points we originally associated to the Northern Arm indeed do not belong to the Eastern Arm and the other way around. The orbit plots of these test are shown in the Appendix, Figure~\ref{im: northeast}.

Overall, all our results agree with previous works on the minispiral \citep[e.g.][]{paumard04, zhao09, zhao10}. In contrast to the other works, we use a Bayesian analysis to get the orbital fits to the gas streamers, which allows us to investigate also the posterior distribution for our result. Further, we include as free parameters the period of the orbits to not only get the best fit spatially but also for the line-of-sight velocity.

\section{\textbf{Conclusions}}\label{sec:Conclusion}

The goal of this paper was to use KMOS data to analyze the kinematics of the ionized gas at the Galactic center and to measure the mass of the black hole, Sgr A$^{\ast}$. We started with the spectra of the ionized hydrogen gas in the central $\sim$ 2 pc. From the Doppler-shifted position of the Br$\gamma$ line, we extracted the line-of-sight velocities of the gas. We also looked at the flux and the FWHM of the Br$\gamma$ line to have a better overview where the gas is more concentrated and to see clearly the features of Sgr A West.
Using these data, we then identified the three main features of Sgr A West: the Northern Arm, Eastern Arm and Western Arc.

Similar to \cite{zhao09} we fitted to these three streamers Keplerian orbits, however we used a Bayesian method for that, which allows us to investigate also the posterior distribution of our results. Moreover, we included the velocity measurements to constrain the orbits. This enabled us to not only find the best fit solution but also see how well constrained the parameters are and if there are degenerate solutions. The results of the maximum likelihood fitting provided all the parameters to characterize the orbits of the streamers in three dimensional space and their velocities. With these orbital parameters we then calculated the mass of Sgr A$^{\ast}$, which lies in focus of the elliptical orbits. 

The results are ($6.4_{-2.5}^{+3.9}$, $14.9_{-10.4}^{+69.4}$, $2.8_{-1.0}^{+3.3}$) $\cdot$ $10^6 M_{\odot}$ for the Northern Arm, Eastern Arm and Western Arc, respectively, and the periods of one complete orbit ($35.3^{+14.0}_{-8.5}$, $17.4^{+31.0}_{-11.6}$,  $79.9^{+32.2}_{-20.8}$) $\cdot 10^3$ yr. 

Overall, our results agree with the known values of the mass of the black hole derived from stellar orbits \citep[e.g.][]{gillessen17, boehle16} within our uncertainties. Moreover, our fits provide a very good description of the three-dimensional geometry of the gas orbits as well as their period around the central black hole.

\section*{Acknowledgements}
We are thankful to the anonymous referee for the useful comments and suggestions. We thank Michele Cappellari and Morgan Fouesneau for their helpful advice and input regarding this project. N.N. gratefully acknowledges support by the Deutsche Forschungsgemeinschaft (DFG, German Research Foundation) -- Project-ID 138713538 -- SFB 881 (``The Milky Way System'', subproject B8).

\newpage
\bibliography{mybib}

\begin{thebibliography}{}

\bibitem[{Boehle} et~al., 2016]{boehle16}
{Boehle}, A., {Ghez}, A.~M., {Sch{\"o}del}, R., {Meyer}, L., {Yelda}, S.,
  {Albers}, S., {Martinez}, G.~D., {Becklin}, E.~E., {Do}, T., {Lu}, J.~R.,
  {Matthews}, K., {Morris}, M.~R., {Sitarski}, B., and {Witzel}, G. (2016).
\newblock {An Improved Distance and Mass Estimate for Sgr A* from a Multistar
  Orbit Analysis}.
\newblock {\em \apj}, 830:17.

\bibitem[{Cappellari} et~al., 2013]{cappellari13}
{Cappellari}, M., {Scott}, N., {Alatalo}, K., {Blitz}, L., {Bois}, M.,
  {Bournaud}, F., {Bureau}, M., {Crocker}, A.~F., {Davies}, R.~L., {Davis},
  T.~A., {de Zeeuw}, P.~T., {Duc}, P.-A., {Emsellem}, E., {Khochfar}, S.,
  {Krajnovi{\'c}}, D., {Kuntschner}, H., {McDermid}, R.~M., {Morganti}, R.,
  {Naab}, T., {Oosterloo}, T., {Sarzi}, M., {Serra}, P., {Weijmans}, A.-M., and
  {Young}, L.~M. (2013).
\newblock {The ATLAS$^{3D}$ project - XV. Benchmark for early-type galaxies
  scaling relations from 260 dynamical models: mass-to-light ratio, dark
  matter, Fundamental Plane and Mass Plane}.
\newblock {\em \mnras}, 432:1709--1741.

\bibitem[{Christopher} et~al., 2005]{Christopher2005}
{Christopher}, M.~H., {Scoville}, N.~Z., {Stolovy}, S.~R., and {Yun}, M.~S.
  (2005).
\newblock {HCN and HCO$^{+}$ Observations of the Galactic Circumnuclear Disk}.
\newblock {\em \apj}, 622(1):346--365.

\bibitem[{Davis} et~al., 2013]{Davis13}
{Davis}, T.~A., {Bureau}, M., {Cappellari}, M., {Sarzi}, M., and {Blitz}, L.
  (2013).
\newblock {A black-hole mass measurement from molecular gas kinematics in
  NGC4526}.
\newblock {\em \nat}, 494(7437):328--330.

\bibitem[{Ekers} et~al., 1983]{Ekers83}
{Ekers}, R.~D., {van Gorkom}, J.~H., {Schwarz}, U.~J., and {Goss}, W.~M.
  (1983).
\newblock {The radio structure of SGR A}.
\newblock {\em \aap}, 122:143--150.

\bibitem[{Feldmeier-Krause} et~al., 2017a]{Feldmeier-Krause_2017}
{Feldmeier-Krause}, A., {Kerzendorf}, W., {Neumayer}, N., {Sch{\"o}del}, R.,
  {Nogueras-Lara}, F., {Do}, T., {de Zeeuw}, P.~T., and {Kuntschner}, H.
  (2017a).
\newblock {KMOS view of the Galactic Centre - II. Metallicity distribution of
  late-type stars}.
\newblock {\em \mnras}, 464(1):194--209.

\bibitem[{Feldmeier-Krause} et~al., 2015]{feldmeier15}
{Feldmeier-Krause}, A., {Neumayer}, N., {Sch{\"o}del}, R., {Seth}, A.,
  {Hilker}, M., {de Zeeuw}, P.~T., {Kuntschner}, H., {Walcher}, C.~J.,
  {L{\"u}tzgendorf}, N., and {Kissler-Patig}, M. (2015).
\newblock {KMOS view of the Galactic centre. I. Young stars are centrally
  concentrated}.
\newblock {\em \aap}, 584:A2.

\bibitem[{Feldmeier-Krause} et~al., 2017b]{Feldmeier-Krause2017}
{Feldmeier-Krause}, A., {Zhu}, L., {Neumayer}, N., {van de Ven}, G., {de
  Zeeuw}, P.~T., and {Sch{\"o}del}, R. (2017b).
\newblock {Triaxial orbit-based modelling of the Milky Way Nuclear Star
  Cluster}.
\newblock {\em \mnras}, 466(4):4040--4052.

\bibitem[{Fritz} et~al., 2011]{Fritz11}
{Fritz}, T.~K., {Gillessen}, S., {Dodds-Eden}, K., {Lutz}, D., {Genzel}, R.,
  {Raab}, W., {Ott}, T., {Pfuhl}, O., {Eisenhauer}, F., and {Yusef-Zadeh}, F.
  (2011).
\newblock {Line Derived Infrared Extinction toward the Galactic Center}.
\newblock {\em \apj}, 737:73.

\bibitem[{Genzel} et~al., 2010]{genzel10}
{Genzel}, R., {Eisenhauer}, F., and {Gillessen}, S. (2010).
\newblock {The Galactic Center massive black hole and nuclear star cluster}.
\newblock {\em Reviews of Modern Physics}, 82:3121--3195.

\bibitem[{Genzel} et~al., 1994]{genzel94}
{Genzel}, R., {Hollenbach}, D., and {Townes}, C.~H. (1994).
\newblock {The nucleus of our Galaxy}.
\newblock {\em Reports on Progress in Physics}, 57:417--479.

\bibitem[{Ghez} et~al., 2008]{ghez08}
{Ghez}, A.~M., {Salim}, S., {Weinberg}, N.~N., {Lu}, J.~R., {Do}, T., {Dunn},
  J.~K., {Matthews}, K., {Morris}, M.~R., {Yelda}, S., {Becklin}, E.~E.,
  {Kremenek}, T., {Milosavljevic}, M., and {Naiman}, J. (2008).
\newblock {Measuring Distance and Properties of the Milky Way's Central
  Supermassive Black Hole with Stellar Orbits}.
\newblock {\em \apj}, 689:1044--1062.

\bibitem[{Gillessen} et~al., 2017]{gillessen17}
{Gillessen}, S., {Plewa}, P.~M., {Eisenhauer}, F., {Sari}, R., {Waisberg}, I.,
  {Habibi}, M., {Pfuhl}, O., {George}, E., {Dexter}, J., {von Fellenberg}, S.,
  {Ott}, T., and {Genzel}, R. (2017).
\newblock {An Update on Monitoring Stellar Orbits in the Galactic Center}.
\newblock {\em \apj}, 837:30.

\bibitem[{Gravity Collaboration} et~al., 2019]{abuter19}
{Gravity Collaboration}, {Abuter}, R., {Amorim}, A., {Baub{\"o}ck}, M.,
  {Berger}, J.~P., {Bonnet}, H., {Brand ner}, W., {Cl{\'e}net}, Y., {Coud{\'e}
  Du Foresto}, V., {de Zeeuw}, P.~T., {Dexter}, J., {Duvert}, G., {Eckart}, A.,
  {Eisenhauer}, F., {F{\"o}rster Schreiber}, N.~M., {Garcia}, P., {Gao}, F.,
  {Gendron}, E., {Genzel}, R., {Gerhard}, O., {Gillessen}, S., {Habibi}, M.,
  {Haubois}, X., {Henning}, T., {Hippler}, S., {Horrobin}, M.,
  {Jim{\'e}nez-Rosales}, A., {Jocou}, L., {Kervella}, P., {Lacour}, S.,
  {Lapeyr{\`e}re}, V., {Le Bouquin}, J.~B., {L{\'e}na}, P., {Ott}, T.,
  {Paumard}, T., {Perraut}, K., {Perrin}, G., {Pfuhl}, O., {Rabien}, S.,
  {Rodriguez Coira}, G., {Rousset}, G., {Scheithauer}, S., {Sternberg}, A.,
  {Straub}, O., {Straubmeier}, C., {Sturm}, E., {Tacconi}, L.~J., {Vincent},
  F., {von Fellenberg}, S., {Waisberg}, I., {Widmann}, F., {Wieprecht}, E.,
  {Wiezorrek}, E., {Woillez}, J., and {Yazici}, S. (2019).
\newblock {A geometric distance measurement to the Galactic center black hole
  with 0.3\% uncertainty}.
\newblock {\em \aap}, 625:L10.

\bibitem[Haario et~al., 2001]{haario01}
Haario, H., Saksman, E., and Tamminen, J. (2001).
\newblock An adaptive metropolis algorithm.
\newblock {\em Bernoulli}, 7(2):223--242.

\bibitem[{Karttunen} et~al., 2003]{karttunen}
{Karttunen}, H., {Kroeger}, P., {Oja}, H., {Poutanen}, M., and {Donner}, K.~J.
  (2003).
\newblock {\em {Fundamental astronomy}}.

\bibitem[{Lo} and {Claussen}, 1983]{Lo83}
{Lo}, K.~Y. and {Claussen}, M.~J. (1983).
\newblock {High-resolution observations of ionized gas in central 3 parsecs of
  the Galaxy - Possible evidence for infall}.
\newblock {\em \nat}, 306:647--651.

\bibitem[{Mills} et~al., 2013]{Mills2013}
{Mills}, E.~A.~C., {G{\"u}sten}, R., {Requena-Torres}, M.~A., and {Morris},
  M.~R. (2013).
\newblock {The Excitation of HCN and HCO$^{+}$ in the Galactic Center
  Circumnuclear Disk}.
\newblock {\em \apj}, 779(1):47.

\bibitem[{Mitzkus} et~al., 2017]{mitzkus17}
{Mitzkus}, M., {Cappellari}, M., and {Walcher}, C.~J. (2017).
\newblock {Dominant dark matter and a counter-rotating disc: MUSE view of the
  low-luminosity S0 galaxy NGC 5102}.
\newblock {\em \mnras}, 464:4789--4806.

\bibitem[{Nogueras-Lara} et~al., 2018]{nogueras-lara18}
{Nogueras-Lara}, F., {Gallego-Calvente}, A.~T., {Dong}, H., {Gallego-Cano}, E.,
  {Girard}, J.~H.~V., {Hilker}, M., {de Zeeuw}, P.~T., {Feldmeier-Krause}, A.,
  {Nishiyama}, S., {Najarro}, F., {Neumayer}, N., and {Sch{\"o}del}, R. (2018).
\newblock {GALACTICNUCLEUS: A high angular resolution JHK$_{s}$ imaging survey
  of the Galactic centre. I. Methodology, performance, and near-infrared
  extinction towards the Galactic centre}.
\newblock {\em Astronomy and Astrophysics}, 610:A83.

\bibitem[{Paumard} et~al., 2004]{paumard04}
{Paumard}, T., {Maillard}, J.-P., and {Morris}, M. (2004).
\newblock {Kinematic and structural analysis of the Minispiral in the Galactic
  Center from BEAR spectro-imagery}.
\newblock {\em \aap}, 426:81--96.

\bibitem[{Requena-Torres} et~al., 2012]{Requena-Torres2012}
{Requena-Torres}, M.~A., {G{\"u}sten}, R., {Wei{\ss}}, A., {Harris}, A.~I.,
  {Mart{\'\i}n-Pintado}, J., {Stutzki}, J., {Klein}, B., {Heyminck}, S., and
  {Risacher}, C. (2012).
\newblock {GREAT confirms transient nature of the circum-nuclear disk}.
\newblock {\em \aap}, 542:L21.

\bibitem[{Roberts} and {Goss}, 1993]{roberts93}
{Roberts}, D.~A. and {Goss}, W.~M. (1993).
\newblock {Multiconfiguration VLA H92-alpha observations of Sagittarius A West
  at 1 arcsecond resolution}.
\newblock {\em \apjs}, 86:133--152.

\bibitem[{Roberts} et~al., 1996]{roberts96}
{Roberts}, D.~A., {Yusef-Zadeh}, F., and {Goss}, W.~M. (1996).
\newblock {Kinematics of the Ionized Gas in Sagittarius A West: Mass Estimates
  of the Inner 0.13 Parsecs of the Galaxy}.
\newblock {\em \apj}, 459:627.

\bibitem[{van den Bosch} and {van de Ven}, 2009]{vandenBosch2009}
{van den Bosch}, R.~C.~E. and {van de Ven}, G. (2009).
\newblock Recovering the intrinsic shape of early-type galaxies.
\newblock {\em \mnras}, 398:1117--1128.

\bibitem[{Walsh} et~al., 2013]{Walsh13}
{Walsh}, J.~L., {Barth}, A.~J., {Ho}, L.~C., and {Sarzi}, M. (2013).
\newblock {The M87 Black Hole Mass from Gas-dynamical Models of Space Telescope
  Imaging Spectrograph Observations}.
\newblock {\em \apj}, 770(2):86.

\bibitem[{Wollman} et~al., 1977]{wollman77}
{Wollman}, E.~R., {Geballe}, T.~R., {Lacy}, J.~H., {Townes}, C.~H., and {Rank},
  D.~M. (1977).
\newblock {Ne II 12.8 micron emission from the galactic center. II.}
\newblock {\em The Astrophysical Journal}, 218:L103--L107.

\bibitem[{Zhao} et~al., 2010]{zhao10}
{Zhao}, J.-H., {Blundell}, R., {Moran}, J.~M., {Downes}, D., {Schuster}, K.~F.,
  and {Marrone}, D.~P. (2010).
\newblock {The High-density Ionized Gas in the Central Parsec of the Galaxy}.
\newblock {\em \apj}, 723:1097--1109.

\bibitem[{Zhao} et~al., 2009]{zhao09}
{Zhao}, J.-H., {Morris}, M.~R., {Goss}, W.~M., and {An}, T. (2009).
\newblock {Dynamics of Ionized Gas at the Galactic Center: Very Large Array
  Observations of the Three-dimensional Velocity Field and Location of the
  Ionized Streams in Sagittarius A West}.
\newblock {\em \apj}, 699:186--214.

\end{thebibliography}
\bibliographystyle{apalike}

\appendix

\begin{figure}[!ht]
    \begin{minipage}{0.48\linewidth}
      \includegraphics[width=1\columnwidth]{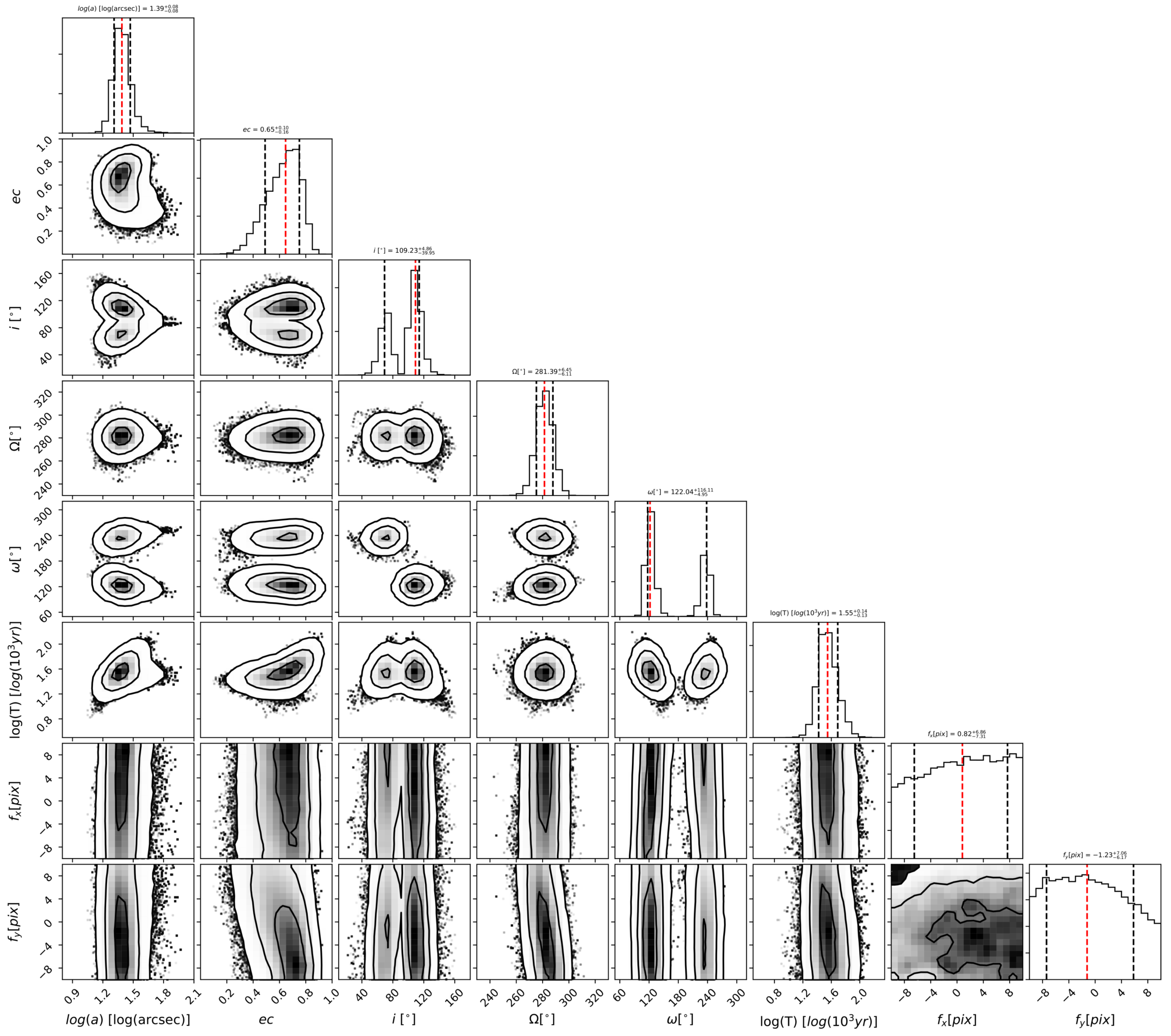}
\caption{The probability distribution function of the orbital parameters for the fit of the Northern Arm.}\label{im:MCMC n}
    \end{minipage}
    \hfill
    \begin{minipage}{0.48\linewidth}
      \includegraphics[width=1\columnwidth]{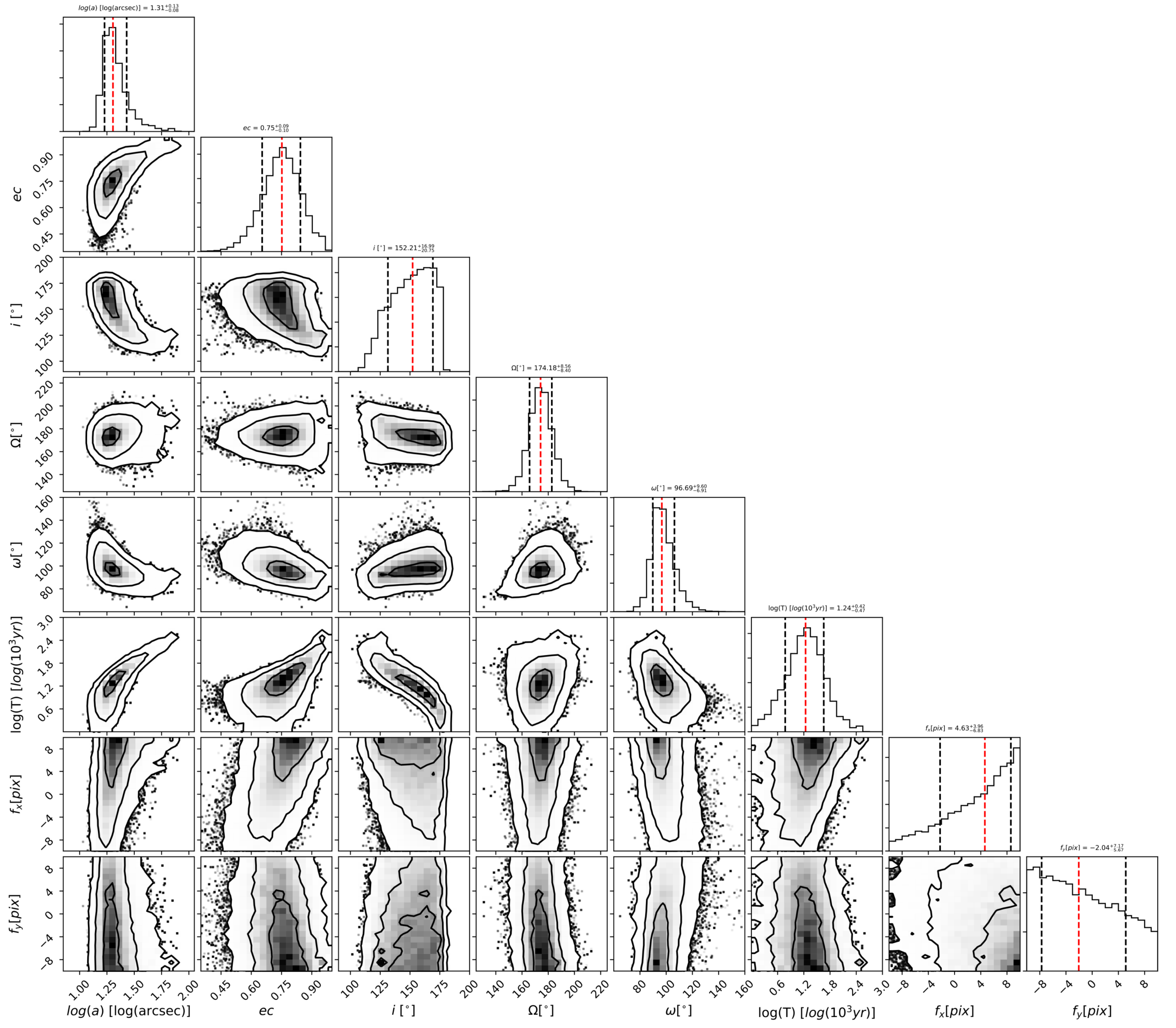}
\caption{The probability distribution function of the orbital parameters for the fit of the Eastern Arm.}\label{im:MCMC e}
    \end{minipage}
    \end{figure}
 \begin{figure}[!ht]  
 \centering
\includegraphics[width=0.48\columnwidth]{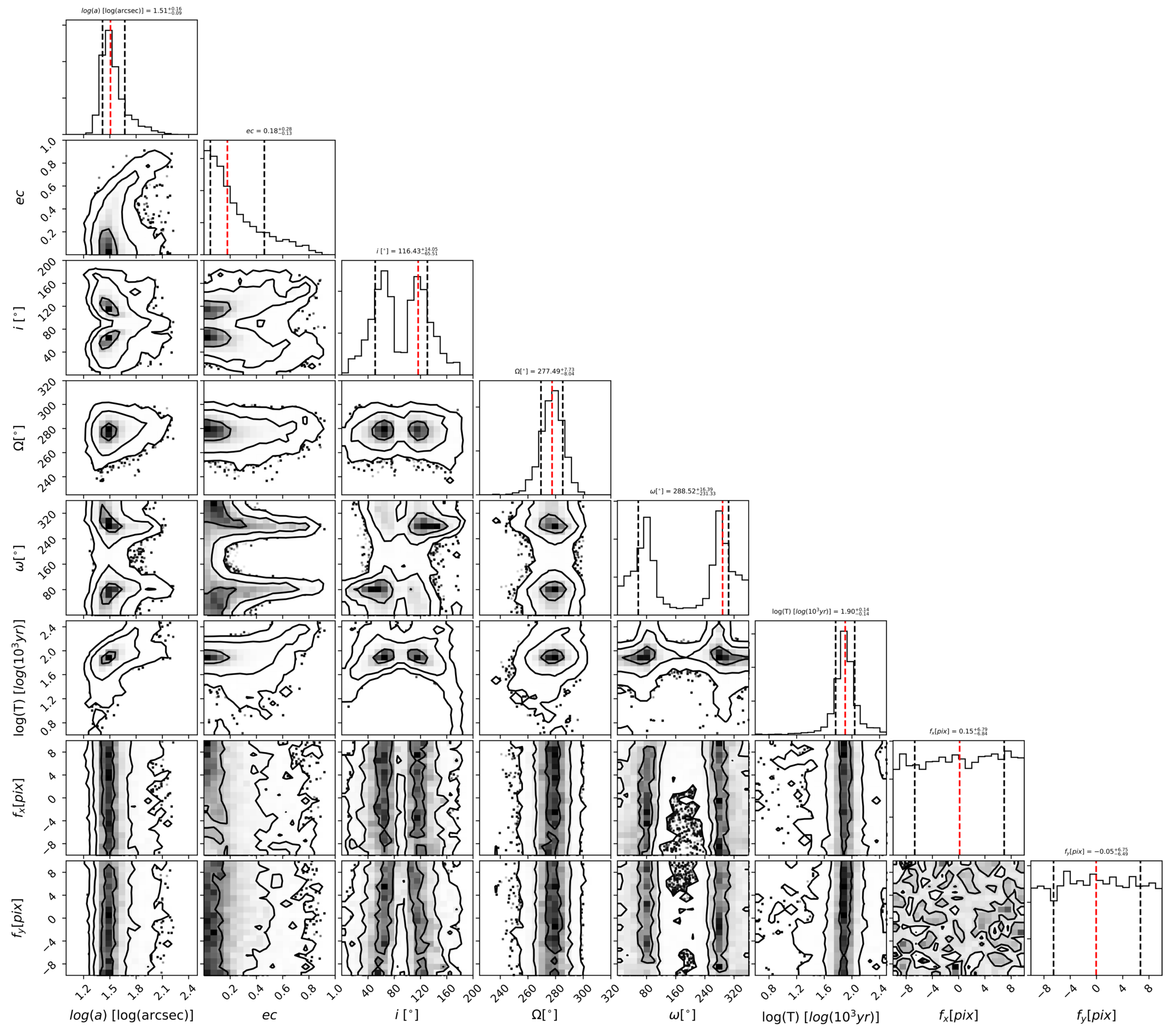}
\caption{The probability distribution function of the orbital parameters for the fit of the Western Arc. The red dashed line is the best-fitting solution and the black dashed lines are the 1$\sigma$ interval for the whole range (for both solutions), this is the same also for the previous two plots.}\label{im:MCMC w}
\end{figure}

\begin{splitdeluxetable}{cccccc|ccccccBcccccc}
\tablecaption{Data points\label{table:data} for Figure~\ref{im:vel}, \ref{im:time} and \ref{im:total plot}}
\tablewidth{0pt} 
\tabletypesize{\scriptsize}
\tablehead{
\multicolumn{6}{c}{Northern Arm} & \multicolumn{6}{c}{Eastern Arm} & \multicolumn{6}{c}{Western Arc}\\
\colhead{$x$} & \colhead{$\Delta x_{\rm err}$} & \colhead{$y$} & \colhead{$\Delta y_{\rm err}$} & \colhead{$v$} & \colhead{$\Delta v_{\rm err}$} & \colhead{$x$} & \colhead{$\Delta x_{\rm err}$} & \colhead{$y$} & \colhead{$\Delta y_{\rm err}$} & \colhead{$v$} & \colhead{$\Delta v_{\rm err}$} &\colhead{$x$} & \colhead{$\Delta x_{\rm err}$} & \colhead{$y$} & \colhead{$\Delta y_{\rm err}$} & \colhead{$v$} & \colhead{$\Delta v_{\rm err}$}\\
\colhead{[pc]} & \colhead{pc]} & \colhead{[pc]} & \colhead{[pc]} & \colhead{[km\,s$^{-1}$]} & \colhead{[km\,s$^{-1}$]} & \colhead{[pc]} & \colhead{[pc]} & \colhead{[pc]} & \colhead{[pc]} & \colhead{[km\,s$^{-1}$]} & \colhead{[km\,s$^{-1}$]} &\colhead{[pc]} & \colhead{[pc]} & \colhead{[pc]} & \colhead{[pc]} & \colhead{[km\,s$^{-1}$]} & \colhead{[km\,s$^{-1}$]}\\
}
\startdata
0.4028 & 0.0793 & -0.0732 & 0.0793 & 89.81 & 5.499 &-0.0426 & 0.0793 & -0.4107 & 0.0793 & 122.6 & 8.096&-1.282 & 0.1189 & -0.4806 & 0.1189 & -64.92 & 10.08 \\
0.06367 & 0.0793 & -0.1529 & 0.0793 & -4.909 & 2.427 &-0.07237 & 0.0793 & -0.4908 & 0.0793 & 155 & 14.08&-1.209 & 0.1189 & -0.233 & 0.1189 & -56.4 & 18.35\\
0.4494 & 0.0793 & -0.0324 & 0.0793 & 81.54 & 0.53 &-0.1155 & 0.0793 & -0.3491 & 0.0793 & 161.8 & 4.811&-0.954 & 0.1189 & 0.05088 & 0.1189 & -98.55 & 17.25\\
0.535 & 0.0793 & 0.03315 & 0.0793 & 137.1 & 5.559 &0.114 & 0.0793 & -0.6528 & 0.0793 & 142 & 9.113&-1.055 & 0.1189 & -0.05469 & 0.1189 & -95.87 & 21.32\\
0.629 & 0.0793 & 0.08304 & 0.0793 & 118 & 10.1 &0.1279 & 0.0793 & -0.5731 & 0.0793 & 135.6 & 9.48&-0.6914 & 0.1189 & 0.3429 & 0.1189 & -58.05 & 18.22\\
0.7147 & 0.0793 & 0.1486 & 0.0793 & 95.38 & 296.1 &0.05931 & 0.0793 & -0.6859 & 0.0793 & 129 & 7.984&-0.5331 & 0.1189 & 0.3547 & 0.1189 & -46.75 & 6.456\\
0.2369 & 0.0793 & -0.101 & 0.0793 & 43.61 & 4.725 &0.2706 & 0.0793 & -0.5697 & 0.0793 & 123.6 & 9.561&-0.1791 & 0.1189 & 0.4744 & 0.1189 & -80.11 & 42.67\\
0.158 & 0.0793 & -0.1188 & 0.0793 & 36.96 & 2.391 &0.2662 & 0.0793 & -0.7125 & 0.0793 & 143.2 & 13.19&0.6682 & 0.1189 & 0.5347 & 0.1189 & 32.75 & 13.43\\
0.3473 & 0.0793 & -0.07456 & 0.0793 & 80.02 & 3.572 &0.5034 & 0.0793 & -0.683 & 0.0793 & 99.51 & 6.737&0.5158 & 0.1189 & 0.6024 & 0.1189 & 25.8 & 14.1\\
0.6762 & 0.0793 & 0.1001 & 0.0793 & 114.6 & 10.24 &0.4319 & 0.0793 & -0.6768 & 0.0793 & 97.19 & 5.279&0.7119 & 0.1189 & 0.3692 & 0.1189 & 79.32 & 6.082\\
-0.168 & 0.0793 & 0.2381 & 0.0793 & -64.81 & 3.645 &0.7403 & 0.0793 & -0.6375 & 0.0793 & 74.1 & 10.2&0.7185 & 0.1189 & 0.4249 & 0.1189 & 85.84 & 15.55\\
0.8852 & 0.0793 & 0.3114 & 0.0793 & 82.97 & 12.89 &0.7544 & 0.0793 & -0.5658 & 0.0793 & 76.67 & 6.624&-0.267 & 0.1189 & 0.5039 & 0.1189 & -26.45 & 34.01\\
0.9799 & 0.0793 & 0.3296 & 0.0793 & 101.8 & 15.08 &0.8497 & 0.0793 & -0.5714 & 0.0793 & 88.35 & 9.712&-1.325 & 0.1189 & -0.005785 & 0.1189 & -89.35 & 21.86\\
1.116 & 0.0793 & 0.5867 & 0.0793 & 86.38 & 21.19 &0.03204 & 0.0793 & -0.5438 & 0.0793 & 126.1 & 10.63&-0.9291 & 0.1189 & 0.3291 & 0.1189 & -49.67 & 4.514\\
-0.1771 & 0.0793 & 0.2854 & 0.0793 & -84.65 & 3.972 &-0.188 & 0.0793 & -0.3033 & 0.0793 & 189.3 & 8.028&-0.9488 & 0.1189 & 0.1621 & 0.1189 & -42.51 & 8.565\\
-0.1548 & 0.0793 & 0.3494 & 0.0793 & -78.7 & 12.3 &-0.1103 & 0.0793 & -0.2379 & 0.0793 & 225 & 24.91&-0.6475 & 0.1189 & 0.4946 & 0.1189 & -19.72 & 19.15\\
-0.1873 & 0.0793 & 0.3804 & 0.0793 & -99.68 & 9.42 &-0.04399 & 0.0793 & 0.2952 & 0.0793 & -176.9 & 4.19&0.5328 & 0.1189 & 0.5552 & 0.1189 & 39.58 & 11.23\\
-0.2339 & 0.0793 & 0.01436 & 0.0793 & -242.5 & 21.86 &-0.08304 & 0.0793 & 0.2704 & 0.0793 & -134.7 & 2.594&0.3931 & 0.1189 & 0.4328 & 0.1189 & 26.1 & 114.8\\
-0.1187 & 0.0793 & -0.1573 & 0.0793 & 58.64 & 247.2 &-0.1561 & 0.0793 & 0.01481 & 0.0793 & -236.1 & 16.8&  \nodata &   \nodata &   \nodata &   \nodata &   \nodata &   \nodata\\
0.8082 & 0.0793 & 0.2143 & 0.0793 & 103.5 & 10.46 &-0.137 & 0.0793 & -0.1196 & 0.0793 & -83.92 & 25.2&  \nodata &   \nodata &   \nodata &   \nodata &   \nodata &   \nodata\\
1.144 & 0.0793 & 0.754 & 0.0793 & 69.62 & 19.16 &-0.1499 & 0.0793 & 0.08635 & 0.0793 & -151.2 & 93.32&  \nodata &   \nodata &   \nodata &   \nodata &   \nodata &   \nodata\\
  \nodata &   \nodata &   \nodata &   \nodata &   \nodata &   \nodata &1.063 & 0.0793 & -0.5186 & 0.0793 & 73.94 & 8.5&  \nodata &   \nodata &   \nodata &   \nodata &   \nodata &   \nodata\\
  \nodata &   \nodata &   \nodata &   \nodata &   \nodata &   \nodata &1.195 & 0.0793 & -0.436 & 0.0793 & 70.97 & 8.788&  \nodata &   \nodata &   \nodata &   \nodata &   \nodata &   \nodata\\
  \nodata &   \nodata &   \nodata &   \nodata &   \nodata &   \nodata &0.02581 & 0.0793 & 0.3603 & 0.0793 & -117.4 & 17.55&  \nodata &   \nodata &   \nodata &   \nodata &   \nodata & \nodata\\
\enddata
\end{splitdeluxetable}

\begin{figure*}[ht]
 \begin{minipage}{0.5\textwidth}
 \includegraphics[width=1\columnwidth]{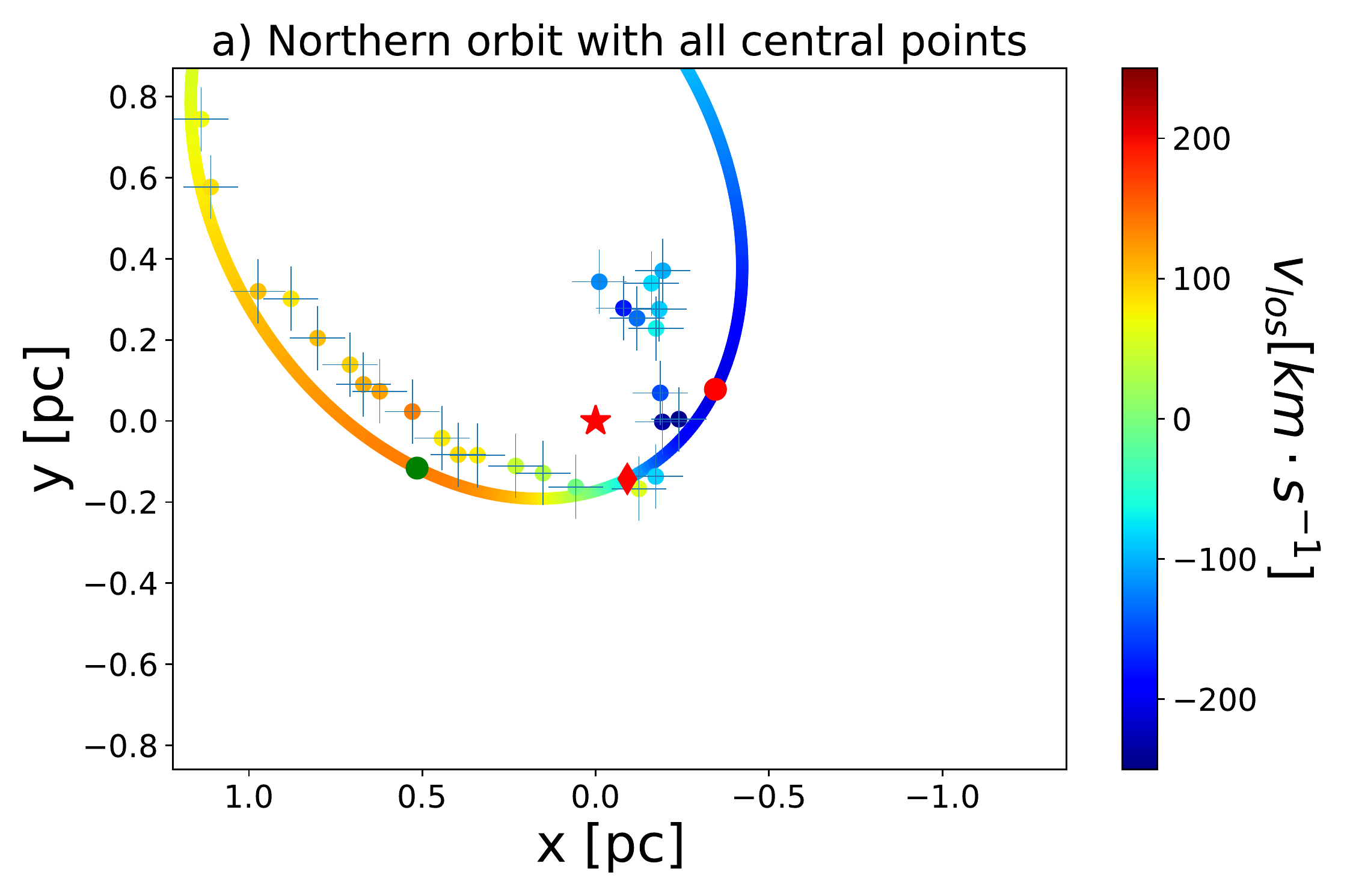}
 \end{minipage}
 \begin{minipage}{0.5\textwidth}
\includegraphics[width=1\columnwidth]{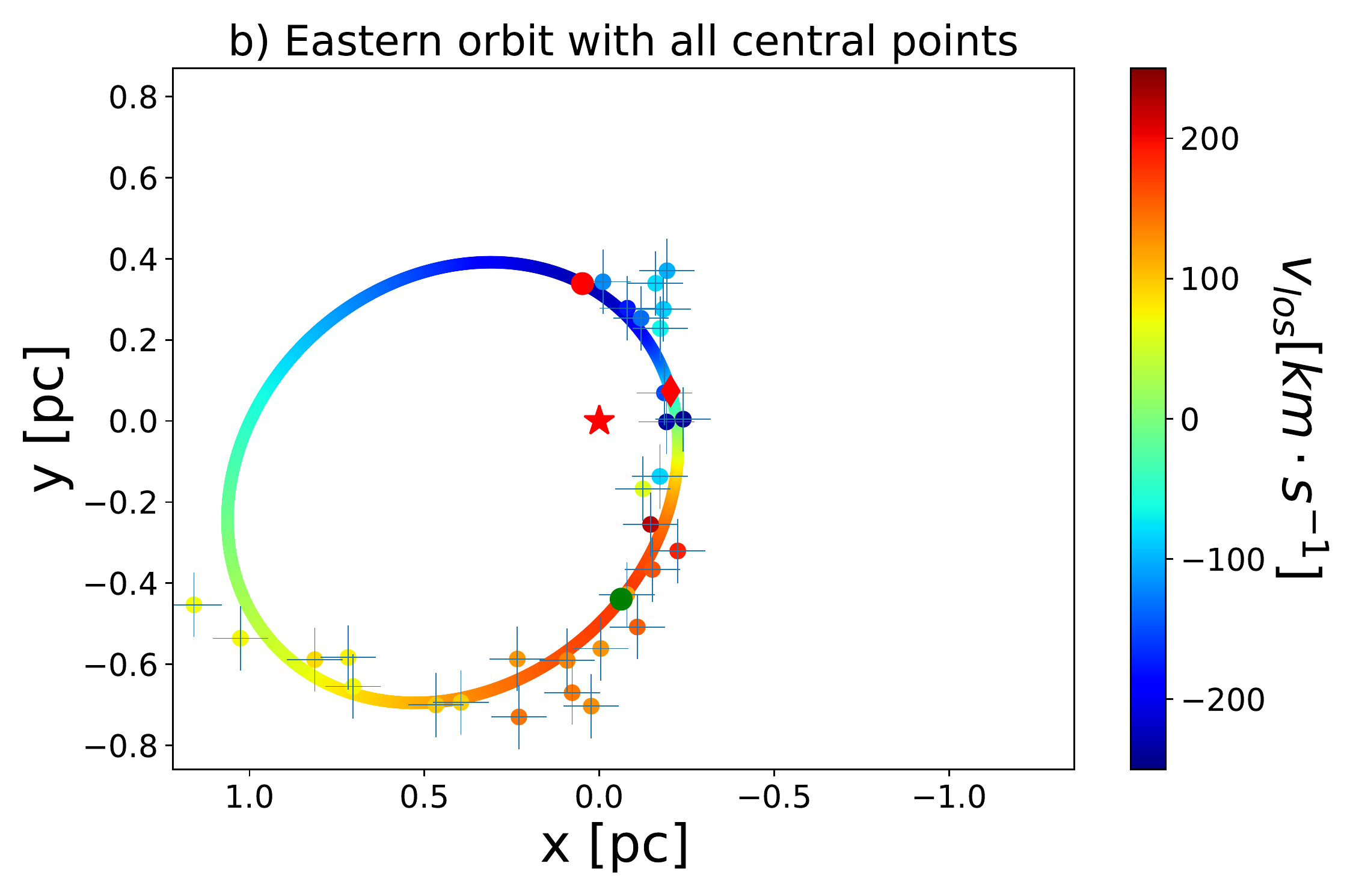}
\end{minipage}
\begin{minipage}{0.5\textwidth}
\centering
\includegraphics[width=1\columnwidth]{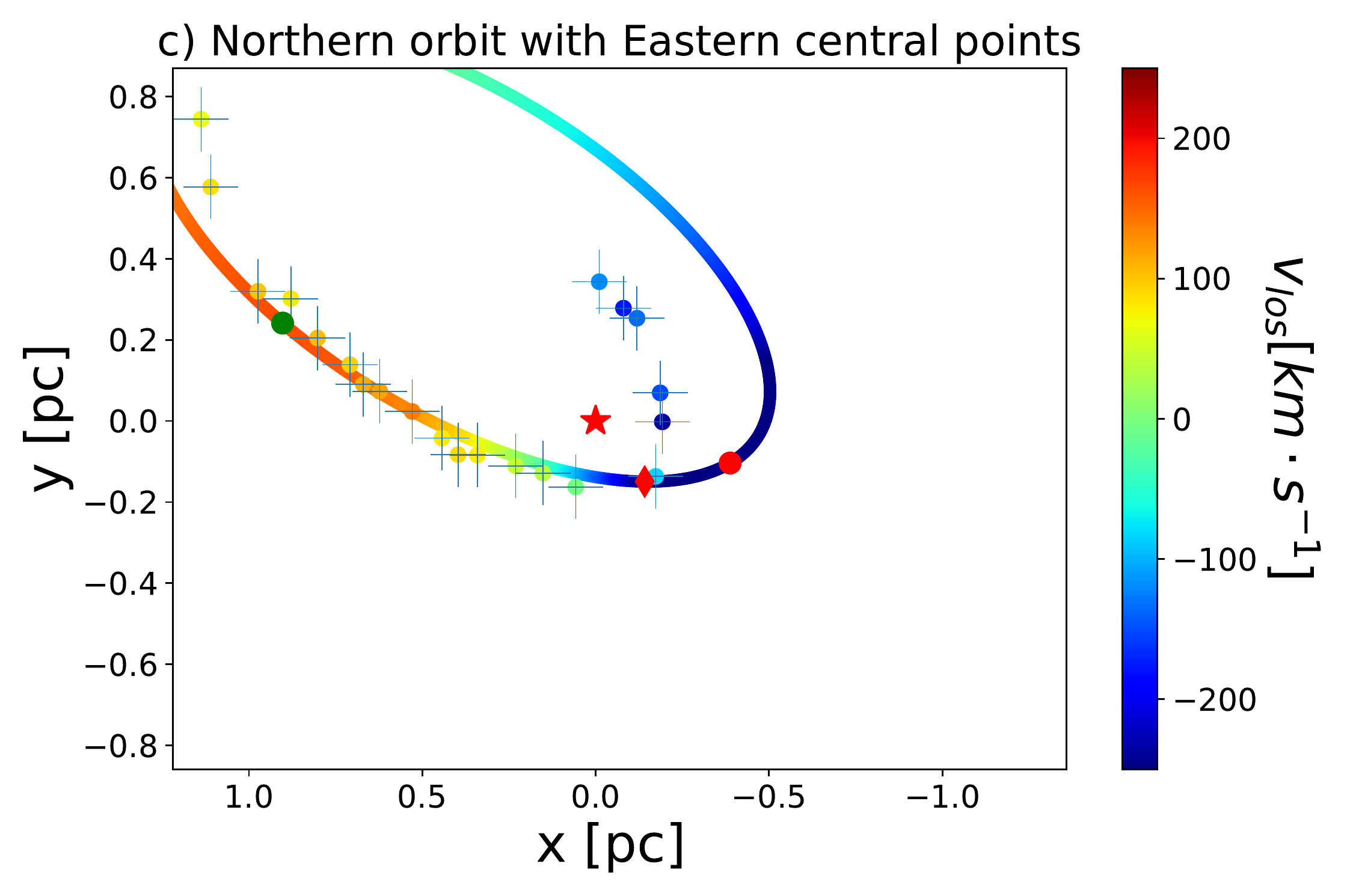}
\end{minipage}
\begin{minipage}{0.5\textwidth}
\centering
\includegraphics[width=1\columnwidth]{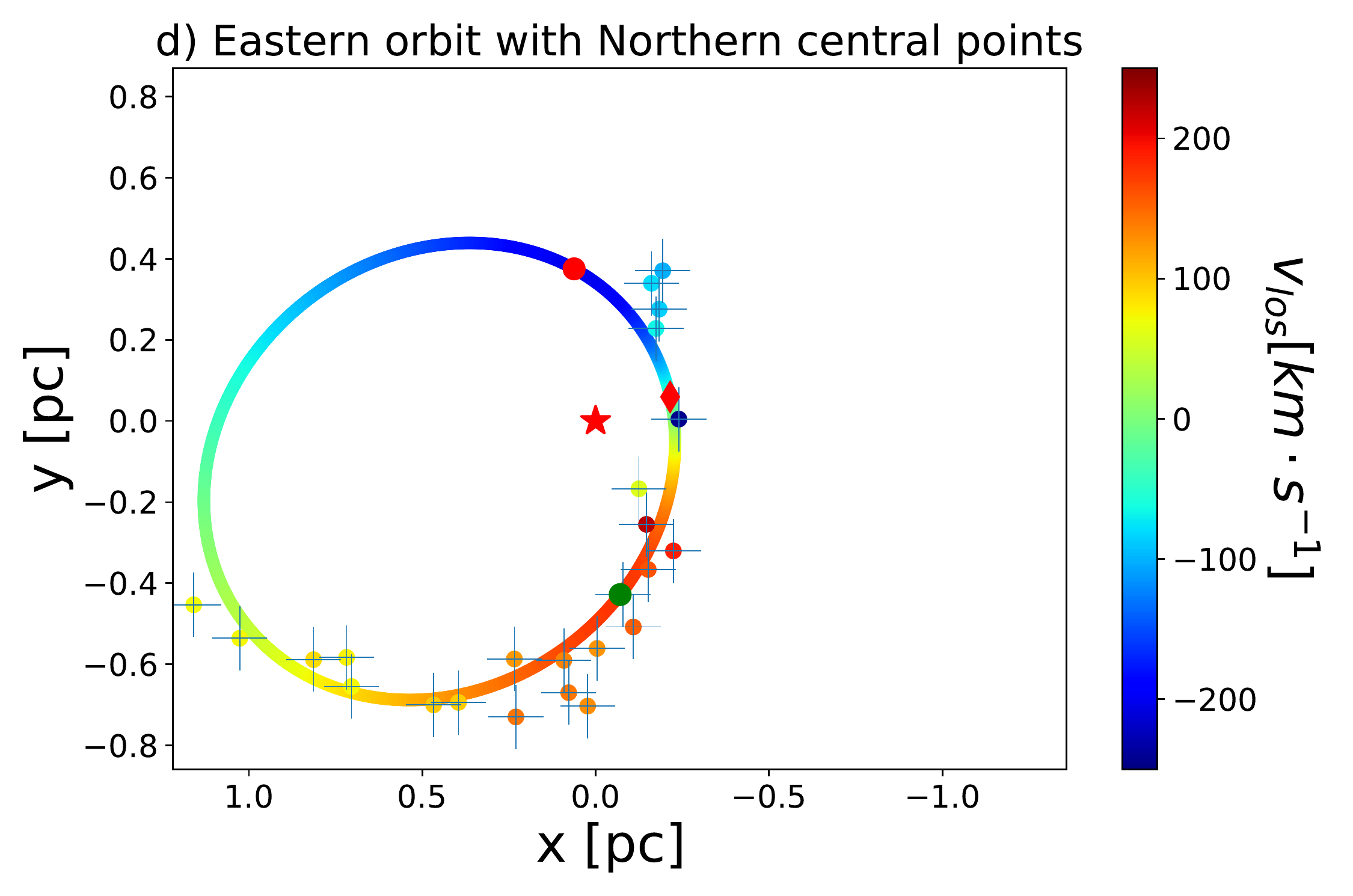}
\end{minipage}
\caption{Different associations of the central points to the Northern and Eastern Arm. A comparison to Fig.~\ref{im:vel} shows that the associations presented here do not work as well as the original ones. a) The Northern Arm with all central points of the original Northern and Eastern orbit, b) the Eastern Arm with all central points of the original Northern and Eastern orbit, c) the Northern Arm with only the central points of the original Eastern Arm and d) the Eastern Arm with only central points of the original Northern Arm. }\label{im: northeast}
\end{figure*}

\end{document}